\documentclass[a4paper,11pt]{article}
\pdfoutput=1 % if your are submitting a pdflatex (i.e. if you have
\usepackage{jcappub} % for details on the use of the package, please
\usepackage[T1]{fontenc} % if needed
\usepackage{graphicx}
\usepackage{amsmath}
\usepackage{scalerel}

\def\apj{ApJ}

\title {\boldmath The long road to the Green Valley: Tracing the
  evolution of the Green Valley galaxies in the EAGLE simulation}

\author[a]{Apashanka Das} 

\author[a]{and Biswajit Pandey}

 \affiliation[a]{Department of Physics, Visva-Bharati University,
  Santiniketan, 731235, India}
\emailAdd{a.das.cosmo@gmail.com}
\emailAdd{biswap@visva-bharati.ac.in}

\abstract{We study the evolution of the progenitors of
    the present-day Green Valley (GV) galaxies across redshift
    $z=10-0$ using data from the EAGLE simulations. We identify the
    present-day green valley galaxies using entropic thresholding and
    track the evolution of the physical properties of their
    progenitors up to $z=10$. Our study identifies three distinct
    phases in their evolution: (i) an early growth phase ($z=10-6$),
    where progenitors are gas-rich, efficiently form stars, and
    experience AGN feedback regulating star formation in massive
    galaxies, (ii) a transition phase ($z=6-2$), marked by frequent
    interactions and mergers in higher-density environments, driving
    starbursts, depleting gas reservoirs, and strengthening
    correlations between cold gas and halo properties, and (iii) a
    quenching phase ($z=2-0$), dominated by environmental and
    mass-dependent processes that suppress star formation and deplete
    cold gas. Our analysis shows that at $z<1$, environmental factors
    and cold gas depletion dominate quenching, with tighter
    correlations between stellar mass, SFR, and cold gas content. The
    interplay between mass and environmental density during this
    period drives diverse and distinct evolutionary pathways. Our
    analysis shows that majority of the main progenitor branches of
    the present-day GV galaxies entered the green valley at $z<1$. We
    also find that a small fraction ($\sim 5\%$) of the main
    progenitor branches had already crossed the green valley and
    joined the red sequence by $z=0.1$, indicating that some galaxies
    may undergo late-time rejuvenation, that allows them to reenter
    the green valley by the present day. Our findings provide a
    comprehensive view of the mechanisms shaping the GV population
    across cosmic time.}
        
\begin{document}
\maketitle
\flushbottom

%\begin{keywords}
%methods: statistical - data analysis - galaxies: formation - evolution
%- cosmology: large scale structure of the Universe.
%\end{keywords}

\section{Introduction}

The observed galaxy properties exhibits a striking bimodality
\citep{strateva, blanton03, bell1, balogh, baldry04a}, a phenomenon
that has captivated astronomers for the last few decades. This
bimodality becomes evident when galaxies are plotted on a
colour-magnitude or colour-stellar mass diagram, where they distinctly
separate into two groups: the blue cloud and the red sequence. The
blue cloud consists of star-forming galaxies characterized by ongoing
star formation and the presence of young, massive stars. In contrast,
the red sequence consists of quiescent galaxies with older stellar
populations, whose redder colours indicate minimal or no recent star
formation. This dichotomy has profound implications for understanding
the processes that drive galaxy evolution over cosmic time.

Deciphering the origins and evolution of this bimodality is crucial
for unraveling the mechanisms behind galaxy formation and
evolution. Semi-analytic models have been developed to replicate
the observed bimodality and provide insights into the underlying
physical processes \citep{menci, driver, cattaneo1, cattaneo2,
  cameron, trayford16, nelson, correa19}. Observations reveal that the
colour bimodality of galaxies persists up to a redshift of $z=1-2$
\citep{bell04b, weiner05, kriek08, brammer09}. It has been found that
the number of massive red galaxies at a fixed stellar mass has
steadily increased since $z \sim 1$ \citep{bell04b, faber07}. This
growth suggests that galaxies in the blue cloud migrate to the red
sequence through the quenching of star formation. The red stellar mass
may also increase via mergers of already quenched, less massive
galaxies. Furthermore, the cosmic star formation rate has declined
sharply between $z=1$ to $z=0$ \citep{madau96}. These trends highlight
significant changes in galaxy properties in recent cosmic history,
which have likely played a pivotal role in shaping the observed
bimodality. Collectively, these observations suggest that galaxies
undergo a transformation from the blue cloud to the red sequence over
time, providing crucial insights into the dynamics of galaxy
evolution.

Between the blue cloud and the red sequence lies the enigmatic ``green
valley'' \citep{wyder07}, a relatively sparse region that represents a
transitional phase in galaxy evolution. Galaxies in the green valley
are thought to be quenching their star formation, transitioning from
the active star-forming state of the blue cloud to the quiescent state
of the red sequence. This phase is significant because it provides a
window into the processes that halt star formation, offering critical
insights into the balance between growth and decline in the cosmic
life-cycle of galaxies.

The distribution of galaxies across the blue cloud, green valley, and
red sequence reflects a dynamic interplay of physical
processes. Star-forming galaxies in the blue cloud often have ample
cold gas reserves fueling star formation. Various environment-driven
processes, including mergers \citep{hopkins08}, harassment
\citep{moore96, moore98}, strangulation \citep{gunn72, balogh00},
starvation \citep{larson80, somerville99, kawata08}, and satellite
quenching \citep{geha12}, can suppress star formation and alter galaxy
structure. Large number of studies using simulations \citep{toomre72,
  barnes96, mihos96, tissera02, cox06, montuori10, lotz11, torrey12,
  hopkins13, renaud14, renaud15, moreno15, moreno21, renaud22, das23}
and observations \citep{larson78, barton00, lambas03, alonso04,
  nikolic04, woods06, woods07, barton07, ellison08, ellison10,
  woods10, patton11, barrera15, thorp22, shah22, das22} show that
interactions between galaxies produce tidal torques which can trigger
starbursts and significantly affect their colour and morphology.

In addition to external factors, internal mechanisms also play a
crucial role in quenching star formation. Processes such as
morphological quenching \citep{martig09}, mass quenching
\citep{birnboim03, keres05, dekel06, gabor10}, angular momentum
quenching \citep{peng20}, and bar quenching \citep{masters10} can lead
to the cessation of star formation. Gas loss, whether through feedback
from supernovae, active galactic nuclei, shock-driven winds
\citep{cox04, murray05, springel05}, or ram pressure stripping
\citep{gunn72}, provides another effective route for quenching.

Once galaxies enter the red sequence, they settle into a quiescent
phase where significant star formation is unlikely without external
disruptions. The journey from the blue cloud through the green valley
to the red sequence can take varying amounts of time, depending on the
internal and external mechanisms driving the transition. Gradual
quenching, such as secular quenching, occurs as galaxies deplete their
gas reserves over time without replenishment, often as a result of
long-term changes in gas inflow and outflow dynamics. This slow
process, sometimes referred to as ``natural aging'', can span $1-2$
Gyr \citep{martig09, larson80, balogh00}. In contrast, rapid quenching
mechanisms, such as mergers and ram pressure stripping, can halt star
formation in less than $1$ Gyr \citep{gunn72, mihos94}. The
transitional green valley phase is therefore critical for
understanding both the timeline and drivers of galaxy
transformation. Does quenching occur rapidly or gradually? Are
external environmental factors more influential than internal
processes? How do the dominant quenching mechanisms evolve with time?
Are certain processes more influential at particular cosmic epochs? By
studying galaxies in the green valley and their progenitors, one can
identify the dominant pathways that shape the bimodal distribution of
galaxies.

Precisely defining the green valley has been a persistent challenge in
studies of galaxy evolution. Broadly, the green valley is
conceptualized as the transitional region between the star-forming
blue cloud and the quiescent red sequence on colour-magnitude or
colour-stellar mass diagrams. Various methodologies have been proposed
to delineate this region. \cite{schawinski14} delineate the green
valley by employing two empirical lines in the colour-stellar mass
diagram.  \cite{coenda18} map the green valley within the (NUV–r)
colour-stellar mass plane using empirical boundaries, analyzing the
characteristics of transitional galaxies across various
environments. \cite{bremer18} classify galaxies into red, blue, and
green categories using three broad colour bins derived from point
densities in the colour-mass plane. \cite{angthopo19} suggest
defining the green valley based on the strength of the $4000\AA$
break, which was later employed for in-depth studies of stellar
populations in green valley galaxies\citep{angthopo20}. \cite{pandey}
utilize a fuzzy set theory approach to classify galaxies as red, blue,
or green within the SDSS dataset. \cite{das21} and \cite{sarkar22}
utilize this classification to examine the properties of green valley
galaxies and red spirals across various
environments. \cite{quilley22} reinterpret the green valley by
linking galaxy morphology to evolutionary trends, defining it through
the average colour of different Hubble types. \cite{noirot22} adopt
the NUVrK colour-colour diagram to separate the galaxies in the blue
cloud, green valley, and red sequence. \cite{estrada23} define the
green valley based on the shape of the log(sSFR) distribution,
investigating the morphological changes of transitional galaxies
observed in the CLEAR survey. \cite{brambila23} use empirical lines
in the SFR-stellar mass plane to define the green valley, exploring
how different environments affect the quenching of these galaxies. The
absence of a universal definition complicates the objective study of
green valley galaxies and their evolution. The boundaries of the green
valley vary depending on the methodology and datasets used.

To address these inconsistencies, \cite{pandey23} propose a
parameter-free method using Otsu’s image segmentation technique to
distinguish the blue cloud and red sequence. However, this approach
does not explicitly identify the green valley. More recently,
\cite{pandey24} introduced a method for separating the green valley
using entropic thresholding \citep{pun81, kapur85}, which is based on
the principle of maximum entropy. This approach provides a natural and
robust definition of the green valley, with boundaries determined
solely by the data. By eliminating the need for arbitrary thresholds,
this method offers a more general and data-driven framework for
studying the green valley.

The journey of galaxies through the ``green valley'' represents a
pivotal phase in their evolution, acting as a bridge between the
star-forming ``blue cloud'' and the quiescent ``red sequence''.
The primary aim of this study is to explore the journey of galaxies
from the star-forming blue cloud to the present-day green valley over
cosmic time. Galaxies are dynamic cosmic ecosystems, where star
formation is fueled by cold gas and regulated by a delicate balance of
internal and external factors. However, this balance can be disrupted,
leading to the cessation of star formation, a process known as
quenching. Understanding the mechanisms behind quenching is vital for
unraveling the evolution of the galaxy population from the early
Universe to the present day. Numerous quenching mechanisms have been
identified, but they often interact in complex ways, with their
significance varying based on type, mass, and environment of
galaxies. These processes rarely act in isolation, instead, galaxies
frequently experience a combination of quenching influences throughout
their lifetimes. The diversity of quenching pathways highlights the
intricate nature of galaxy evolution and underscores the importance of
detailed investigations, such as those conducted with hydrodynamical
simulations, to disentangle the contributions of these
mechanisms. Deciphering when and how these processes operate is
critical for explaining the observed bimodality in galaxy populations
and understanding the intricate evolutionary pathways that lead
galaxies through the green valley.

Studying the progenitors of present-day green valley galaxies across
cosmic time allows us to uncover the physical drivers of star
formation quenching and assess their relative importance at different
epochs. Hydrodynamical simulations, such as the Evolution and Assembly
of Galaxies and their Environments (EAGLE) project \citep{schaye15},
provide a powerful tool for this purpose. These simulations capture
the complex interplay of gravitational dynamics, hydrodynamic
processes, and feedback mechanisms that govern galaxy evolution,
enabling detailed tracking of individual galaxy histories. By tracing
the progenitors of present-day green valley galaxies from high
redshifts ($z=10$) to the present day ($z=0$), we can disentangle the
cumulative effects of various quenching mechanisms. This comprehensive
approach offers a time-resolved perspective on how galaxies gradually
lose their capacity to sustain star formation, shedding light on the
intricate pathways of galaxy transformation.

Te structure of the paper is as follows: Section 2 provides a
description of the data and the methodology, Section 3 discusses the
results, and Section 4 presents the conclusions.

\section{Data and method of analysis}

\subsection{Data}

We utilize data from the EAGLE suite of cosmological hydrodynamical
simulations of galaxy formation \citep{mcalpine16, schaye15,
  crain15}. These simulations model the evolution of both baryonic and
dark matter from a starting redshift of $z=127$ to the present
day. The EAGLE simulations are based on a flat $\Lambda$CDM
cosmological model, with parameters derived from the Planck Mission
results \cite{planck14}. The galaxies are simulated across cosmic time
within comoving cubic volumes, with side lengths ranging from 25 to
100 Mpc.

In this study, we use the reference model denoted as $Ref-L0100N1504$,
which simulates a total of $2 \times 1504^3$ particles within a
comoving cubic box with a side length of 100 Mpc. The initial masses
assigned to the baryonic and dark matter particles in this model are
$1.81 \times 10^6 M_{sun}$ and $9.70 \times 10^6 M_{sun}$,
respectively. We extract the comoving coordinates of the simulated
galaxies, which are identified by the locations of their minimum
gravitational potential. Only genuine galaxies are selected by
applying the criterion $Spurious = 0$. Additionally, we retrieve
various physical properties of these galaxies, such as stellar mass,
star formation rate, cold gas mass, dark matter mass, and $(u-r)$
colour. These properties are measured within a 3D spherical aperture
of comoving radius 30 kpc centered on the galaxy's minimum
gravitational potential. The $(u-r)$ colour represents the difference
in the rest-frame absolute magnitudes of the galaxies in the SDSS $u$-
and $r$-band filters \citep{doi10, trayford15}.  At the present-day
snapshot ($z = 0, Snapnum=28$), colour information is available only
for galaxies with stellar masses $\log(M_{stellar}/M_{sun}) \geq
8.3$. Each galaxy in our sample is resolved by at least $\sim 110$
particles. Combining all this information, we identify a total of
$29754$ galaxies at $z=0$.

At $z=0$, we classify galaxies in our sample as red, blue, or green
based on their stellar mass and $(u-r)$ colour, using the entropic
thresholding technique described in \autoref{subsec:rbg}. The green
galaxies identified in this manner at $z=0$ include all necessary
information, with each galaxy assigned a unique identifier,
$GalaxyID$. We use the $GalaxyID$ of each green galaxy at $z=0$ to
trace its merger history over past time intervals. In the EAGLE
simulation, each galaxy is described by a main progenitor branch,
which tracks its individual evolution from $z\sim 10$ to the present
day, and by other progenitor branches that record its merger history
at different redshifts. The $LastProgID$ associated with each green
galaxy at $z=0$ represents the highest $GalaxyID$ among all its
progenitors across the redshift range from $z\sim10$ to $z=0.1$. Thus,
for a present-day green galaxy with $GalaxyID=i$ and $LastProgID=j$,
querying all galaxies with their $IDs$ between $i < IDs \leq j$ across
all redshifts retrieves all progenitor galaxies associated with this
green galaxy. Following this approach, we identify all progenitor
galaxies for green galaxies at $z=0$. We retrieve the progenitor
galaxies of present-day green galaxies in redshift range
$z=10-0.1$. Our sample contains $9628$ green galaxies at $z=0$,
classified according to the scheme described in \autoref{subsec:rbg},
and a total of $141099$ progenitor galaxies. It is important to note
that only the green-valley progenitors with stellar masses
$\log(M_{stellar}/M_{sun}) \geq 8.3$ are considered in our
analysis. This ensures that all the progenitors are resolved by at
least $110$ particles. This limit in the lower mass is also essential
as the colours are only available for the galaxies above this stellar
mass in EAGLE simulation.

\subsection{Identifying the present-day green valley galaxies using entropic thresholding}
\label{subsec:rbg}
We first classify the galaxies as red, blue at $z=0$ following Otsu's
technique proposed in \citep{pandey23}. The method identifies the
optimal threshold that separates the bimodal $(u-r)$ colour
distribution into two populations, the Blue Cloud (BC) representing
star-forming galaxies with younger stellar populations and the Red
Sequence (RS) representing passive galaxies with older stellar
populations.

The threshold maximizes the inter-class variance and minimizes the
intra-class variance. Mathematically, the intra-class variance
$\sigma^2_{\text{intra}}$ and inter-class variance
$\sigma^2_{\text{inter}}$ are expressed as,
\begin{equation*}
  \sigma^2_{\text{intra}} = P_{\scaleto{BC}{3.5pt}}\sigma^2_{\scaleto{BC}{3.5pt}} + P_{\scaleto{RS}{3.5pt}}\sigma^2_{\scaleto{RS}{3.5pt}},
\end{equation*} 
\begin{equation*}
\sigma^2_{\text{inter}} = P_{\scaleto{BC}{3.5pt}}
P_{\scaleto{RS}{3.5pt}} (\mu_{\scaleto{BC}{3.5pt}} -
\mu_{\scaleto{RS}{3.5pt}})^2,
\end{equation*}
where $P_{\scaleto{BC}{3.5pt}}$ and $P_{\scaleto{RS}{3.5pt}}$ are the
probabilities of class occurrences, and $\mu_{\scaleto{BC}{3.5pt}}$,
$\mu_{\scaleto{RS}{3.5pt}}$, $\sigma^2_{\scaleto{BC}{3.5pt}}$ and
$\sigma^2_{\scaleto{RS}{3.5pt}}$ are the class means and class
variances for the two populations.

The green valley is defined as the transitional region between the
blue cloud and red sequence. The optimal colour threshold in Otsu's
method divides the entire population into blue cloud and red
sequence. However this does not help us to identify the green valley
sandwiched between the blue cloud and the red sequence. We next
classify the galaxies into three distinct classes as red, blue and
green following the technique proposed in \citep{pandey24}, based on
entropic thresholding \cite{kapur85}. We apply entropic thresholding
to identify green valley galaxies at $z = 0$. The entire population is
first divided into three separate classes $(u-r) \leq
\mu_{\scaleto{BC}{3.5pt}}$, $(u-r) \geq \mu_{\scaleto{RS}{3.5pt}}$ and
$\mu_{\scaleto{BC}{3.5pt}} < (u-r) < \mu_{\scaleto{RS}{3.5pt}}$. Here
$\mu_{\scaleto{BC}{3.5pt}},\mu_{\scaleto{RS}{3.5pt}}$ respectively
represents the mean colours associated with the blue cloud and red
sequence identified with the Otsu's method. The populations with
$(u-r) \leq \mu_{\scaleto{BC}{3.5pt}}$ belongs to the blue cloud and
$(u-r) \geq \mu_{\scaleto{RS}{3.5pt}}$ belongs to the red sequence,
and the intermediate region $\mu_{\scaleto{BC}{3.5pt}} < (u-r) <
\mu_{\scaleto{RS}{3.5pt}}$ is represented by the blue, green and red
galaxies. The intermediate region ($\mu_{\scaleto{BC}{3.5pt}} < (u-r)
< \mu_{\scaleto{RS}{3.5pt}}$) is divided into $N$ bins. Entropic
thresholding is employed to partition this region into three distinct
zones belonging to Blue Cloud (BC), Green Valley (GV), and Red
Sequence (RS). This is achieved by maximizing the total entropy
$H_{\text{total}}$ of this region
\begin{equation}
H_{\text{total}} = H_{BC} + H_{GV} + H_{RS}
\end{equation}
Here the entropy associated with each class $H_{\text{BC}}$ (blue
cloud), $H_{GV}$ (green valley), $H_{RS}$ (red sequence)
is given by,
\begin{equation} 
H_{BC}=\log(\sum^{s_1}_{i=1}p_i)-\frac{\sum^{s_1}_{i=1} p_i \log p_i}{\sum^{s_1}_{i=1} p_i}
\end{equation}

\begin{equation}
H_{GV}=\log(\sum^{s_2}_{i=s_1+1}p_i)-\frac{\sum^{s_2}_{i=s_1+1} p_i \log p_i}{\sum^{s_2}_{i=s_1+1} p_i}
\end{equation}

\begin{equation}
H_{RS}=\log(\sum^N_{i=s_2+1}p_i)-\frac{\sum^N_{i=s_2+1} p_i \log p_i}{\sum^N_{i=s_2+1} p_i}
\end{equation}

Here $p_i$ represents the probability of the $i^{th}$ colour bin, and
we choose $N=30$ bins for our analysis. Previous studies indicate that
the boundaries of the green valley defined by entropic thresholding
are not significantly sensitive to the choice of the number of bins
\citep{pandey24}. The thresholds $s_1$ and $s_2$ represent possible
colour values between $\mu_{\scaleto{BC}{3.5pt}}$ and
$\mu_{\scaleto{RS}{3.5pt}}$. To identify these thresholds, we iterate
over all possible combinations of $s_1$ and $s_2$ (with $s_1<s_2$) in
the range $[\mu_{\scaleto{BC}{3.5pt}}, \mu_{\scaleto{RS}{3.5pt}}]$ and
determine the values that maximize the total entropy, $H_{total} =
H_{BC} + H_{GV} + H_{RS}$. The optimal thresholds $s_1$ and $s_2$ thus
obtained partition the population into three distinct classes: the
Blue Cloud ($(u-r) \leq s_1$), the Red Sequence ($(u-r) \geq s_2$),
and the Green Valley ($s_1 < (u-r) < s_2$).

To account for the dependence of galaxy colour on stellar mass, we
separately apply this classification process within independent
stellar mass bins. Within each bin, we first determine the mean
colours $\mu_{\scaleto{BC}{3.5pt}}$ and $\mu_{\scaleto{RS}{3.5pt}}$
for the blue cloud and red sequence using Otsu's technique
\citep{pandey23}. We then separate the Green Valley (GV) by
identifying two thresholds $(s_1,s_2)$ within
$\mu_{\scaleto{BC}{3.5pt}}$ and $\mu_{\scaleto{RS}{3.5pt}}$, using
entropic thresholding \citep{pandey24}. This ensures that the
boundaries of the green valley vary dynamically with stellar mass,
providing a nuanced classification.

\begin{figure*}[htbp!]
\centering
\includegraphics[width=15cm]{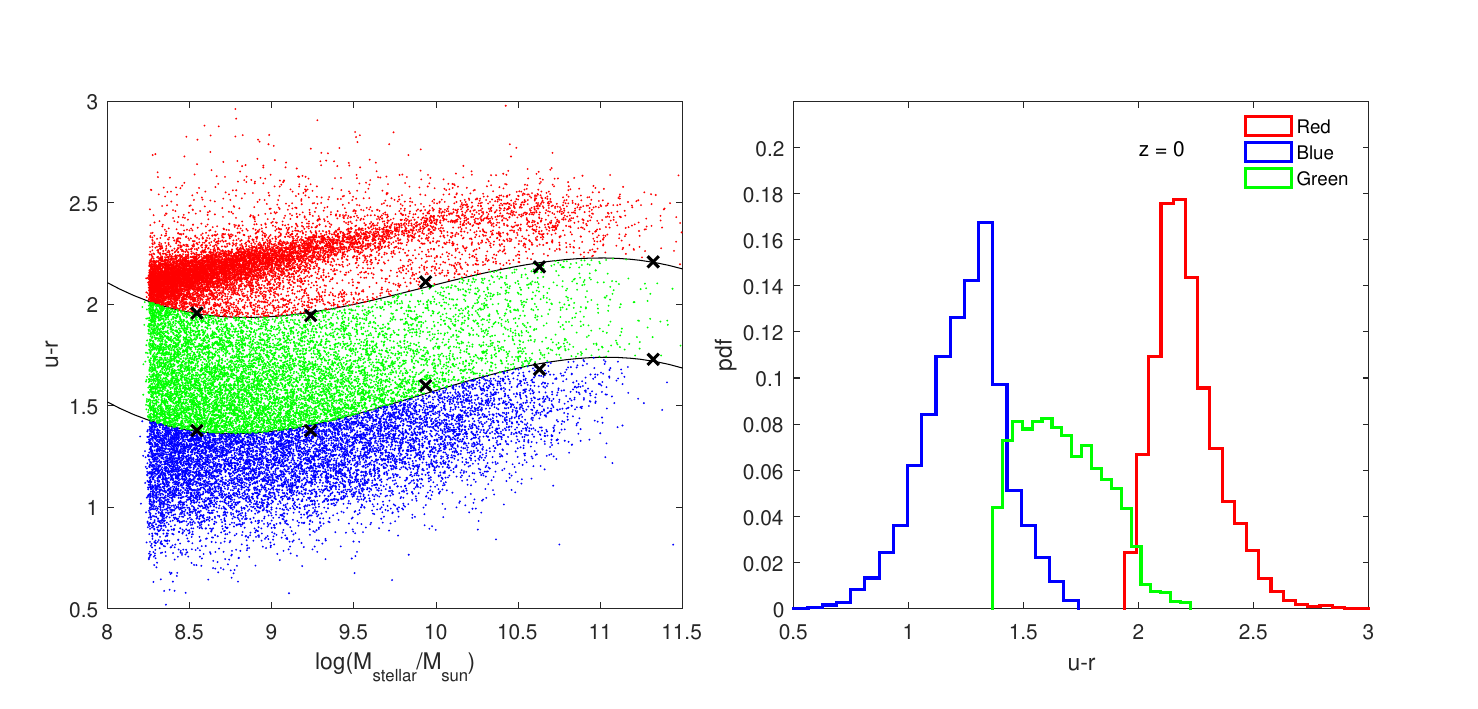}
\caption{The left panel illustrates red, blue, and green galaxies
  classified in the colour-stellar mass plane using the entropic
  thresholding technique. Black cross markers indicate the boundaries
  of the green valley across different stellar mass bins, with the
  solid black line representing a cubic polynomial fit to these
  points. The right panel displays the PDF of $(u-r)$ colour for red,
  blue, and green galaxies, also classified using entropic
  thresholding. Both panels correspond to redshift of $z=0$.}
\label{otsu_fig}
\end{figure*}

The left panel of \autoref{otsu_fig} illustrates this classification,
where the cross points indicate the thresholds $s_1$ and $s_2$ for
each mass bin, and the solid line represents a smooth cubic polynomial
fit to these thresholds in the colour-stellar mass plane. The right
panel of \autoref{otsu_fig} shows the PDF of $(u-r)$ colour for the
red, blue, and green galaxy populations, classified using the entropic
thresholding technique at $z=0$.

\subsection{Quantifying the local environment of green valley galaxies and their progenitors}
We quantify the local environment of present-day green valley galaxies
and their progenitors by estimating their local density using the
method described in \cite{casertano85}. This approach calculates the
local density $\eta_k$ of a galaxy based on the distance to its
$k^{th}$ nearest neighbour as,
\begin{equation}
\eta_k=\frac{k-1}{\frac{4}{3}\pi r^3_k}.
\label{eq:lden}
\end{equation}
Here, $r_k$ denotes the distance to the $k^{th}$ nearest neighbour,
and the denominator represents the volume of a sphere with radius
$r_k$ centered on the galaxy. For our analysis, we calculate the local
density (\autoref{eq:lden}) using the distance to the $5^{th}$ nearest
neighbour ($k=5$). Since the EAGLE simulation is conducted within a
finite volume with well-defined boundaries, the local density for
galaxies near the boundary can be underestimated. This underestimation
occurs because some portion of the measuring spheres can lie outside
the simulation boundary. To address this issue, we exclude the
galaxies from our sample if $r_5 > r_b$, where $r_5$ represents the
distance to the $5^{th}$ nearest neighbour and $r_b$ represents the
distance of the galaxy from the nearest boundary. We consider the
entire snapshot at each redshift for determining the local density for
the green valley galaxies and their progenitors.

\subsection{Identifying AGN activity in green valley galaxies and their progenitors}
The presence of an active galactic nucleus (AGN) significantly impacts
star formation in a galaxy. When a supermassive black hole accretes
material, it converts the gravitational energy of the infalling matter
into radiative energy. This radiative energy is emitted into the
interstellar medium, disrupting the cold gas reservoirs that are
essential for sustaining star formation. The radiation, winds, and
jets from the AGN can expel cold gas or heat it, a process known as
AGN feedback.

Galaxies actively forming stars can transition to quiescence due to
AGN activity, and their emission line flux ratios can be used to
distinguish between star-forming and AGN-active galaxies
\citep{bpt}. AGN-active galaxies are characterized by strong emission
lines compared to other galaxies. In this study, we classify
AGN-active galaxies among the present-day green valley population and
their progenitors in the EAGLE hydrodynamical simulation using a
luminosity threshold, following the approach outlined in
\citep{stuart20}. Galaxies with a bolometric AGN luminosity of
$L_{bol} \geq 10^{43}$ erg/s are classified as AGN-active.

The bolometric AGN luminosity is calculated as $L_{bol}=\epsilon_r
\dot m_{\scaleto{BH}{3.5pt}} c^2$, where $\epsilon_r=0.1$ is the
radiative efficiency of the accretion disk \citep{shakura73}, $\dot
m_{\scaleto{BH}{3.5pt}}$ is the black hole accretion rate, and $c$ is
the speed of light. This formula represents the fraction of the black
hole’s accreting mass that is converted into radiative energy. In
\autoref{fracagn}, we present the fraction of green valley galaxies at
$z=0$ and their progenitors at various redshifts that host an AGN.

\subsection{Identifying green valley galaxies and their progenitors in interacting pairs}
We identify galaxies involved in interactions using the criteria
outlined in \cite{das23a}. A galaxy is classified as interacting if
its first nearest neighbour lies within a distance of $r \leq 200$
kpc, where $r$ represents the three-dimensional comoving separation in
real space between the locations of the minimum gravitational
potential of the galaxies. Using this approach, we determine which
green valley galaxies in our sample at $z=0$ are undergoing
interactions. Similarly, we identify progenitor galaxies that are part
of interacting pairs at various redshifts. It is worthwhile to mention
that we only focus on interacting galaxy pairs with a stellar mass
ratio within the range $1 \leq \frac{M_1}{M_2} \leq 10$. This
selection ensures that the interactions have a significant impact on
both galaxies involved. While galaxy pairs with more extreme mass
ratios, such as $1:1000$, exist in our sample, their interactions are
unlikely to meaningfully affect the more massive galaxy in the pair
and are therefore excluded from this study. We show the fraction of
present-day green valley galaxies and their progenitors involved in
interacting systems in \autoref{interact}. We use the entire snapshot
at each redshift for identifying the green valley galaxies and their
progenitors in interacting pairs.

\section{Results}
In this section, we analyze the physical properties of the green
valley progenitors across the redshift range $10-0.1$ and explore the
physical mechanisms and processes driving their evolution.

\subsection{AGN activity in the progenitors of the present-day green valley galaxies}
\begin{figure*}[htbp!]
\centering
\includegraphics[width=15cm]{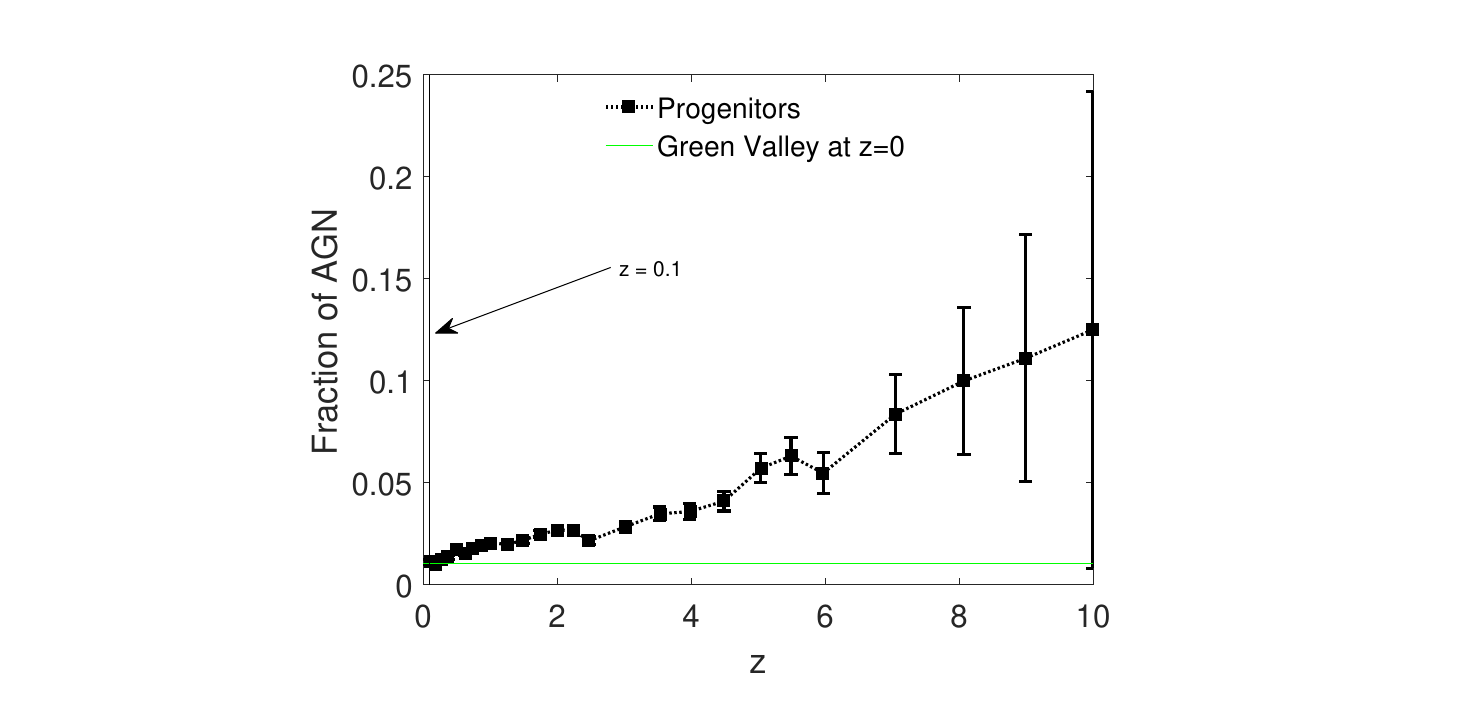}
\caption{This plot shows the fraction of GV progenitors hosting AGN as
  a function of redshift. The corresponding fraction for present-day
  green valley galaxies is indicated by a green horizontal
  line. Binomial 1$\sigma$ error bars are shown for each data point.}
\label{fracagn}
\end{figure*}

To understand the role of AGN feedback in the evolution of GV
progenitors, we calculate the fraction of AGN in these galaxies across
the redshift range $z=10$ to $z=0$. The results, shown in
\autoref{fracagn}, reveal a steady decline in the AGN fraction,
dropping from approximately $12\%$ at $z=10$ to about $2\%$ at
$z=0$. This trend reflects the decreasing significance of AGN feedback
as a driver of galaxy evolution over time.

At low redshifts ($z<2$), the scarcity of cold gas significantly
limited AGN activity, reducing its role in galaxy regulation. At this
stage, quenching became primarily driven by environmental factors,
secular processes, and lingering effects from earlier AGN
feedback. The declining AGN fraction indicates a shift in quenching
mechanisms. AGN feedback is most influential during the early stages
of quenching but plays a diminished role in maintaining quiescence at
later times. As a result, GV progenitors transition from AGN-dominated
regulation at high redshifts to quenching dominated by environmental
effects and cold gas depletion at lower redshifts.

\subsection{Galaxy interactions in the progenitors of the present-day green valley galaxies}
\begin{figure*}[htbp!]
\centering
\includegraphics[width=15cm]{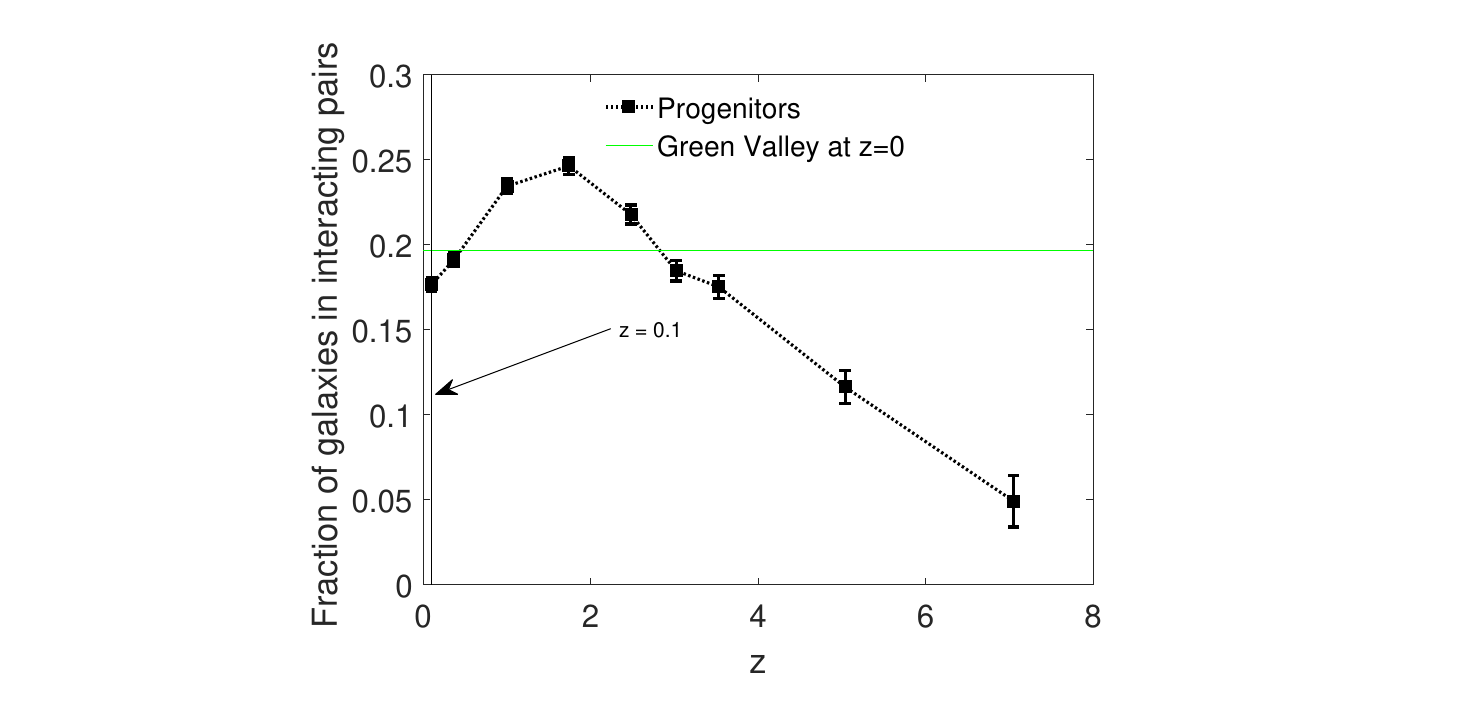}
\caption{The plot shows the fraction of GV progenitors in interacting
  galaxy pairs across different redshifts. A green horizontal line
  indicates the corresponding fraction for present-day green valley
  galaxies. Binomial 1$\sigma$ error bars are displayed for each data
  point.}
\label{interact}
\end{figure*}

We analyze the fraction of interacting galaxies among GV progenitors
across the redshift range $z=10$ to $z=0$, with results shown in
\autoref{interact}. The fraction of interacting galaxies steadily
rises from $\sim 5\%$ at $z=7$ to a peak of $\sim 25 \%$ at $z=2$,
before declining to about $19 \%$ at the present day.

This trend highlights the significant role of galaxy mergers and
interactions in driving galaxy evolution during $2<z<7$. Interactions
during this period are particularly effective at triggering starbursts
and depleting cold gas reservoirs, accelerating the quenching
process. 

The frequency of interactions in low-density regions is expected to
decrease significantly once the universe entered a phase of
accelerated expansion at low redshift. The decline in the interaction
fraction at $z<2$ reflects a shift in quenching mechanisms, with
non-interaction-based processes, including environmental effects,
secular evolution, and feedback mechanisms, playing an increasingly
significant role in shaping the GV population. This evolution
underscores the dynamic interplay of interactions and environment in
the life cycle of GV progenitors.

\subsection{Evolution of star formation rate and cold gas mass in the progenitors of the present-day green valley galaxies}

\begin{figure*}[htbp!]
%\centering
\includegraphics[width=17cm]{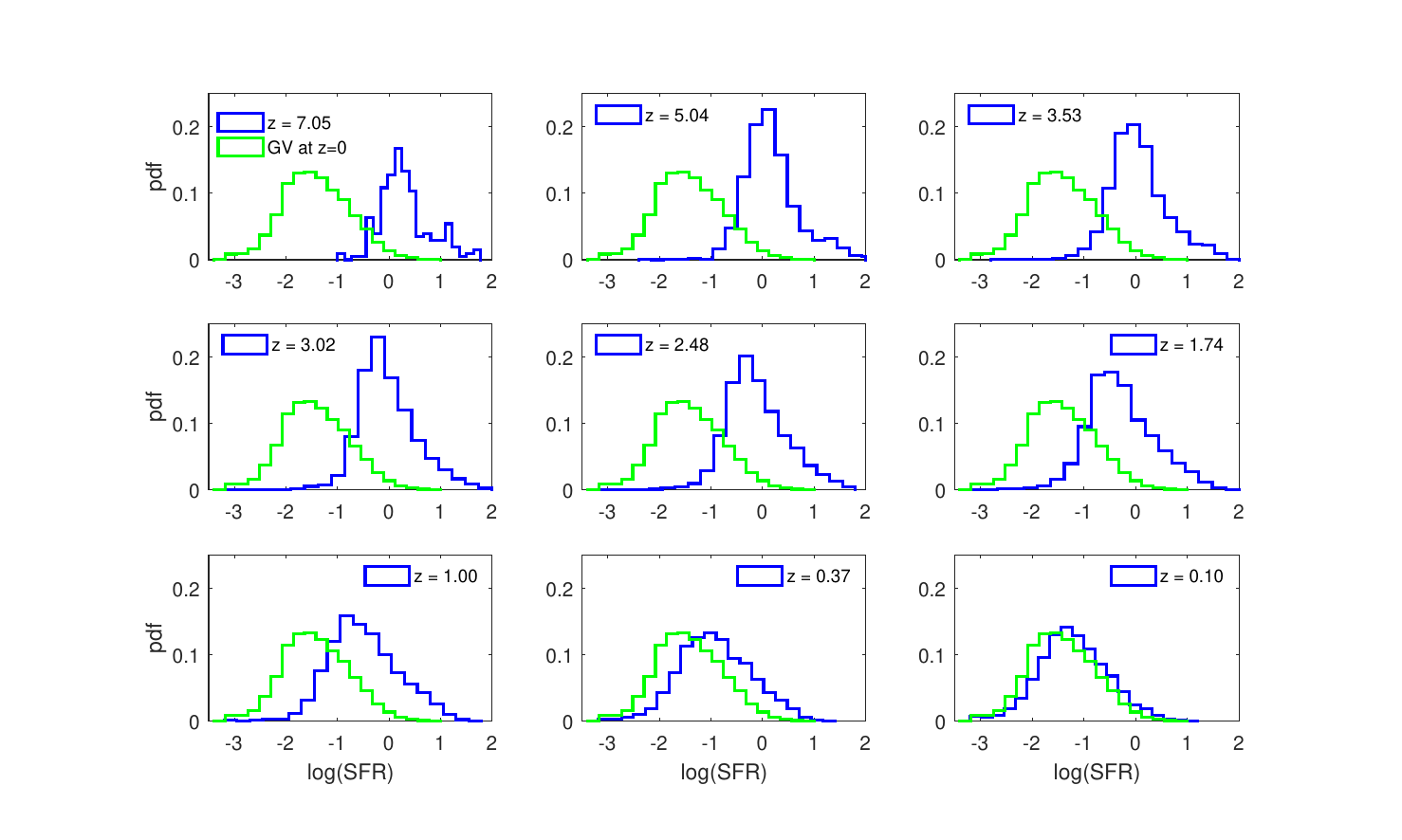}
\caption{This figure presents the probability distribution of SFR in
  GV progenitors across different redshifts, with the SFR distribution
  of present-day green valley galaxies included in each panel for
  comparison.}
\label{sfr}
\end{figure*}

We study the evolution of the star formation rate (SFR) distribution
of GV progenitors as illustrated in the panels of \autoref{sfr}. This
figure compares the SFR distributions of GV progenitors with those of
present-day GV galaxies across various redshifts, as indicated in each
panel. At higher redshifts ($z \geq 5$), GV progenitors predominantly
consist of actively star-forming galaxies. Over time, the peak of the
$\log($SFR$)$ distribution for GV progenitors gradually shifts to lower
values, eventually reaching negative values at redshifts $z \leq
3$. The rate of change in the SFR distribution accelerates at $z < 1$,
where the most dramatic decline occurs, signaling a significant
suppression of star formation in the GV progenitors.

\begin{figure*}[htbp!]
%\centering
\includegraphics[width=17cm]{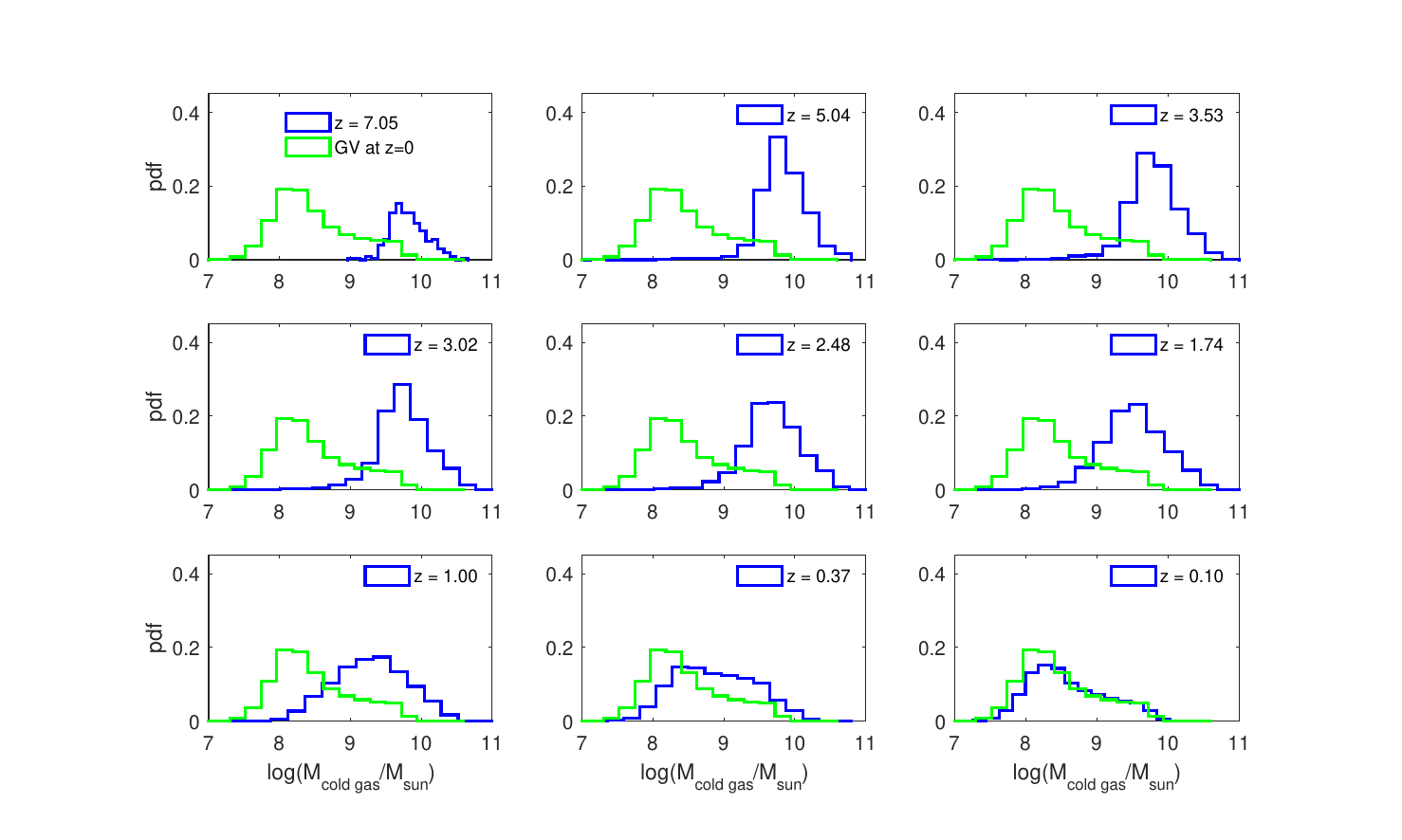}
\caption{This shows the probability distribution of cold gas mass in
  GV progenitors at different redshifts. The results for the
  present-day GV population is shown together in each panel for
  comparison.}
\label{cold}
\end{figure*}

Since cold gas serves as the essential fuel for star formation, we
also track the evolution of cold gas content in GV progenitors over
the same redshift range, as shown in \autoref{cold}. A similar trend
emerges where GV progenitors at high redshifts were gas-rich, with
those at $z \sim 5$ containing, on average, approximately $30$ times
more cold gas than present-day GV galaxies. However, a steady
depletion of cold gas is observed, driven by ongoing star formation
and various gas-destroying processes. This depletion becomes most
pronounced at $z<1$, highlighting a strong correlation between the
suppression of star formation in GV progenitors and the diminishing
availability of cold gas during this period.

We next focus on the stellar mass-SFR relation and halo mass-cold gas
mass relation for the GV progenitors to understand their roles in the
evolution of SFR and cold gas in these galaxies. The mass of galaxies
and their host dark matter halos plays a pivotal role in regulating
star formation. A key framework for studying this relationship is the
galaxy mass-main sequence, which links stellar mass to the SFR. This
correlation reveals that more massive galaxies tend to exhibit higher
SFRs, forming a tight and consistent trend across a broad range of
stellar masses and redshifts. This pattern arises because stellar mass
reflects the capacity of a galaxy to sustain star formation through
ongoing gas accretion and the self-regulation of feedback
processes. Notably, deviations from the mass-main sequence often
signify critical transitions in a galaxy's life cycle, such as
quenching or rejuvenation of star formation. As a result, the
mass-main sequence offers valuable insights into the physical
mechanisms that drive and suppress star formation throughout cosmic
history.

\begin{figure*}[htbp!]
%\centering
\includegraphics[width=17cm]{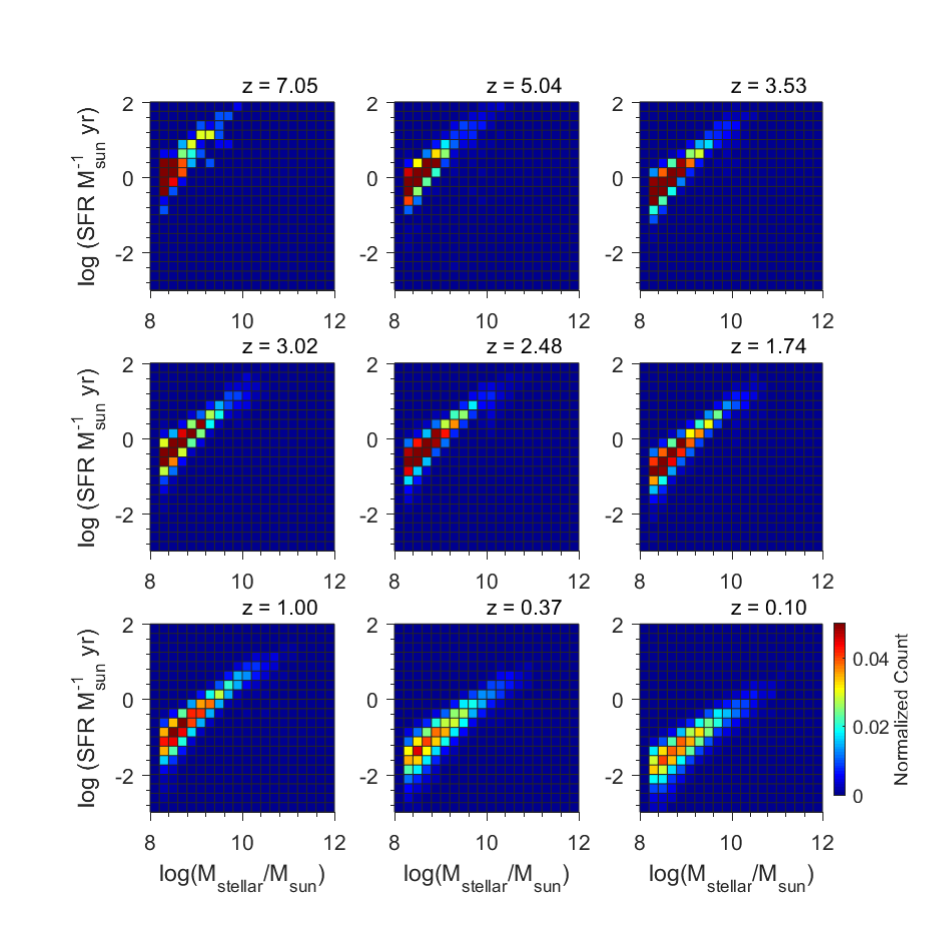}
\caption{This plot shows the joint PDF of stellar mass and SFR of GV
  progenitors at different redshifts. The colour bar at the bottom
  right part of the figure represents the amplitudes of the joint PDF
  of stellar mass and SFR.}
\label{sfrmass}
\end{figure*}

\begin{figure*}[htbp!]
\centering
\includegraphics[width=15cm]{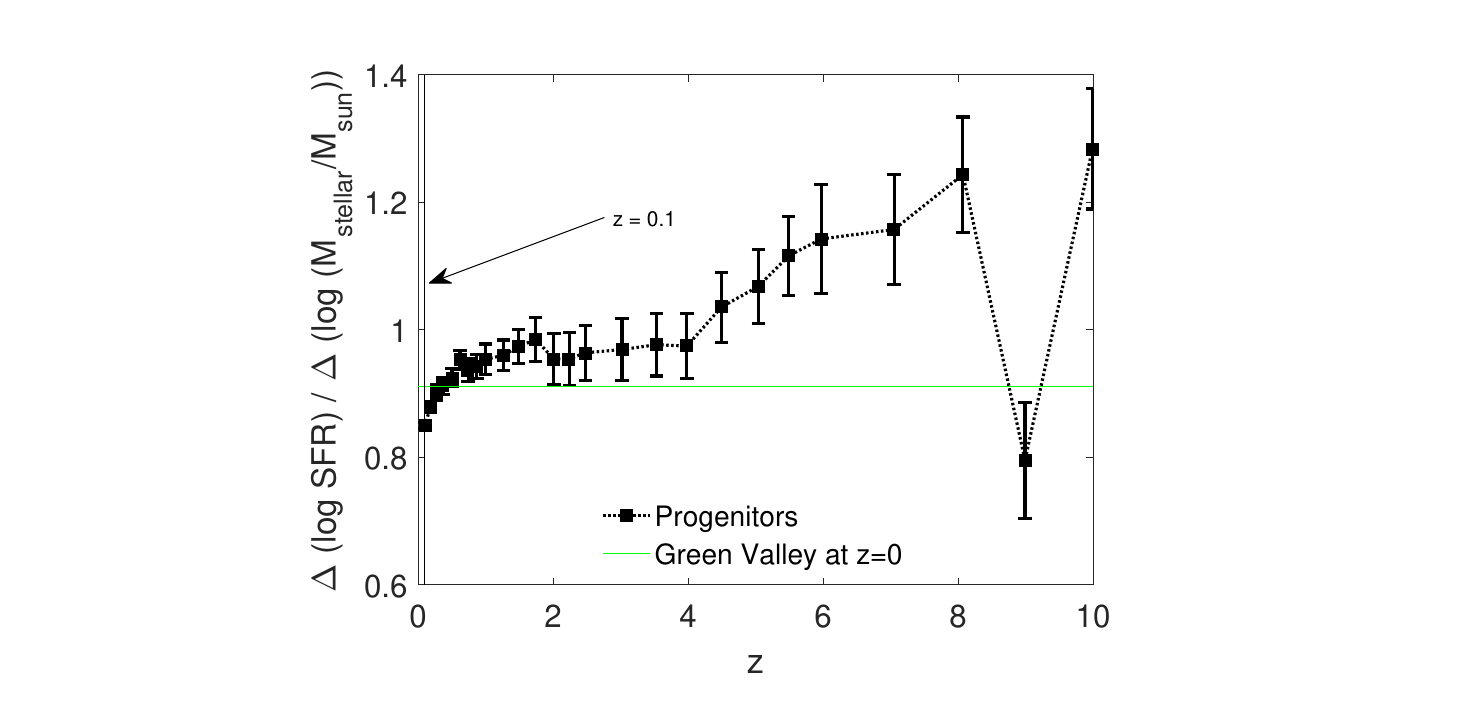}
\caption{This figure shows the slope of the stellar mass-SFR relation
  for GV progenitors as a function of redshift. The slope,
  representing the linear relationship between $\log(M)$ and
  $\log(SFR)$, is calculated using the least-squares fitting
  method. The 1-$\sigma$ error bars obtained through Jackknife
  resampling are shown at each data point.}
\label{sfrmass_slope}
\end{figure*}

\begin{figure*}[htbp!]
%\centering
\includegraphics[width=17cm]{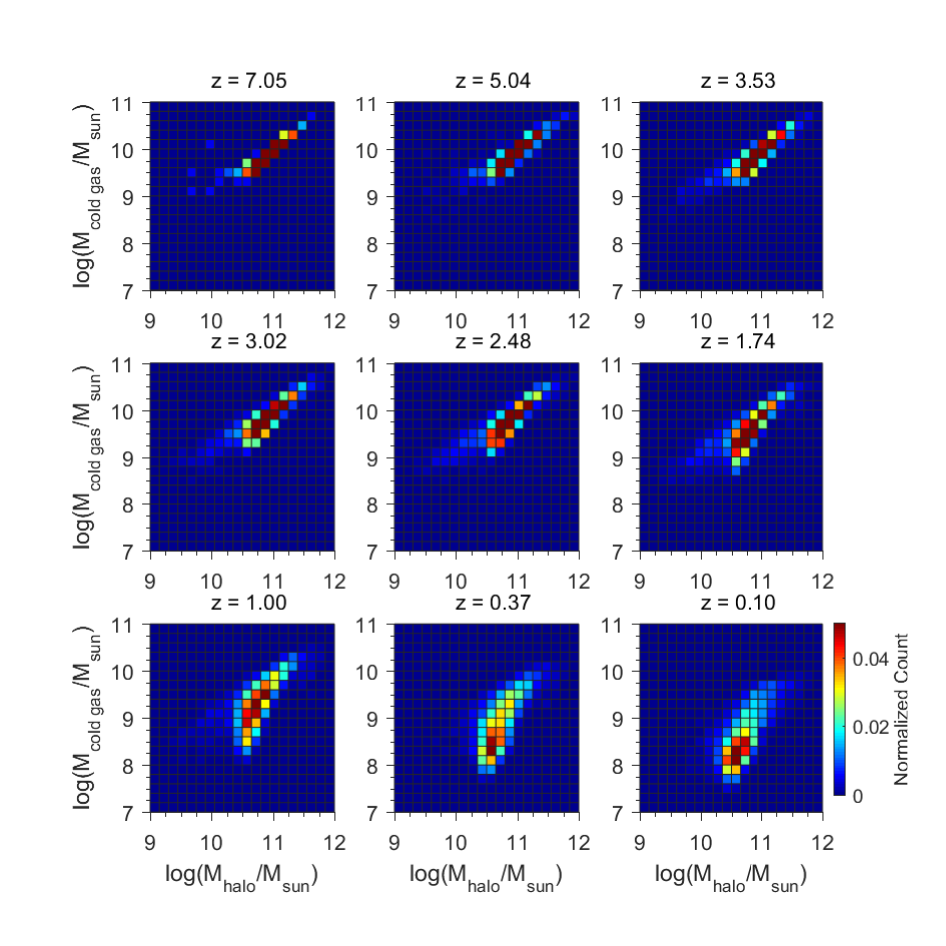}
\caption{This shows the joint PDF of halo mass and cold gas content of
  GV progenitors at different redshifts. The colour bar represents the
  amplitudes of the joint PDF of halo mass and cold gas content.}
\label{coldhalo}
\end{figure*}

\begin{figure*}[htbp!]
\centering
\includegraphics[width=15cm]{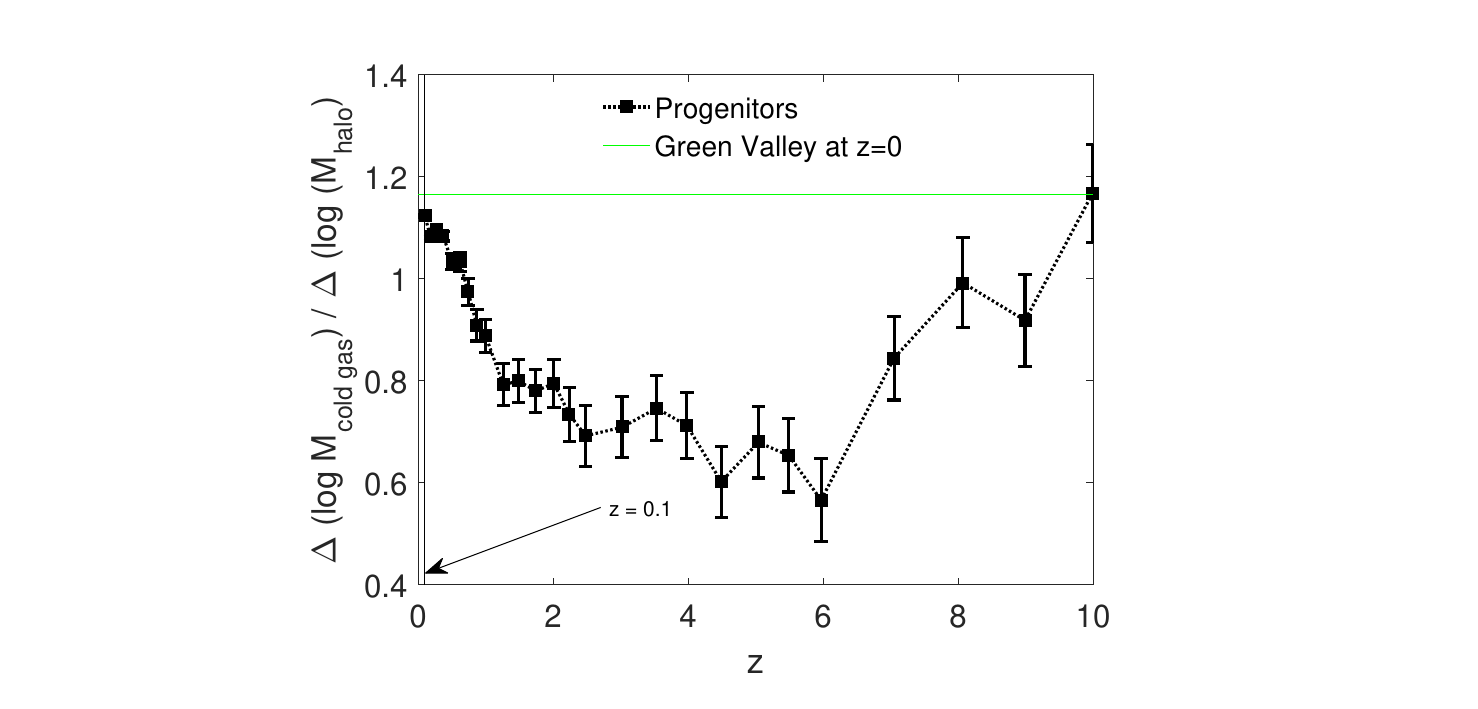}
\caption{This shows the evolution of the slope of the halo mass-cold
  gas mass relation for GV progenitors with redshift. We estimate the
  slope at each redshift using the least-squares fitting method. The
  1-$\sigma$ error bars shown at each data point are obtained through
  Jackknife resampling.}
\label{coldhalo_slope}
\end{figure*}

The panels of \autoref{sfrmass} display the joint PDFs of $\log($SFR$)$
and $\log M_{stellar}$ for GV progenitors across various redshifts, as
indicated in each panel. These visualizations clearly reveal the
evolution of the main-sequence relation of GV progenitors over
time. To quantify this evolution, we calculate the slope of the
main-sequence relation at each redshift using a least-squares fitting
method. The resulting slopes, plotted as a function of redshift in
\autoref{sfrmass_slope}, show distinct trends. The slope decreases
from $z=7$ to $z=4$, stabilizes between $z=4$ and $z=2$, and declines
again from $z=2$ to the present, with the most pronounced changes
occurring at $z<1$.

At high redshifts ($z=7$ to $z=4$), star formation is efficient across
all stellar masses, driven by an abundance of cold gas and frequent
accretion events. As redshift decreases, galaxies of varying masses
may quench their star formation through interactions and mergers with
neighbouring galaxies. In contrast, the secular processes
(e.g. supernova-driven winds) can inhibit star formation in low-mass
progenitors. These processes lead to a gradual flattening of the
main-sequence slope. Between $z=4$ and $z=2$, many progenitors
experience a balance between star formation and feedback regulation,
maintaining a relatively stable slope.

At $z<2$, environmental factors begin to dominate. Higher-mass
progenitors persist longer in the green valley, particularly in
intermediate-density environments, before quenching due to feedback or
environmental influences. In contrast, lower-mass progenitors quench
earlier and more rapidly, as they are less capable of retaining gas
and are more sensitive to feedback effects. By $z<1$, GV progenitors
span a broad range of stellar masses, but most have masses below
$10^{10} M_{sun}$. Consequently, cold gas depletion and environmental
factors drive the quenching of most GV progenitors, resulting in a
dramatic decline in star formation rates at low redshift.

\begin{figure*}[htbp!]
%\centering
\includegraphics[width=17cm]{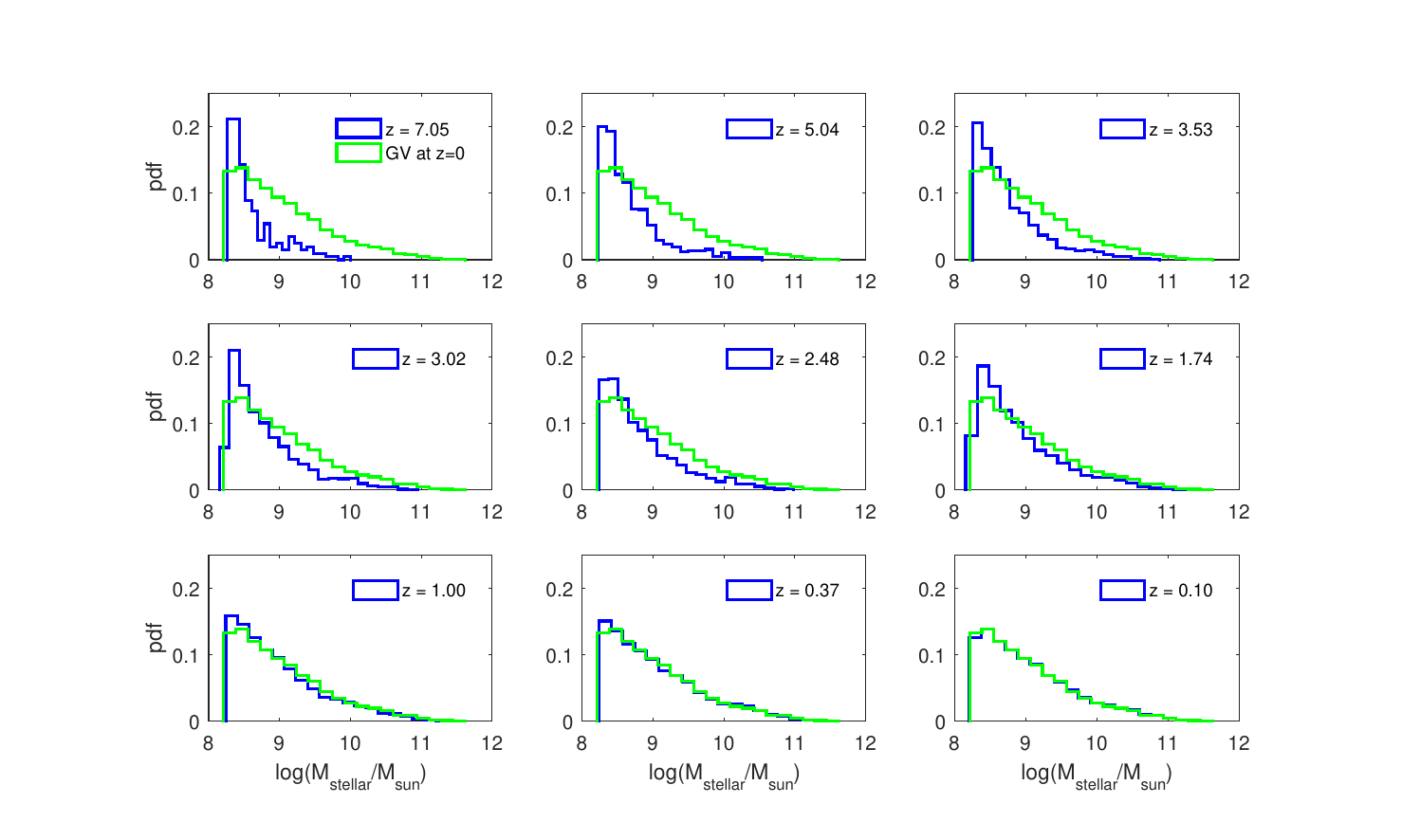}
\caption{This shows the PDF of the stellar mass for GV progenitors at
  different redshift along with that for the present-day green valley
  galaxies.}
\label{pdfmass}
\end{figure*}

\begin{figure*}[htbp!]
%\centering
\includegraphics[width=17cm]{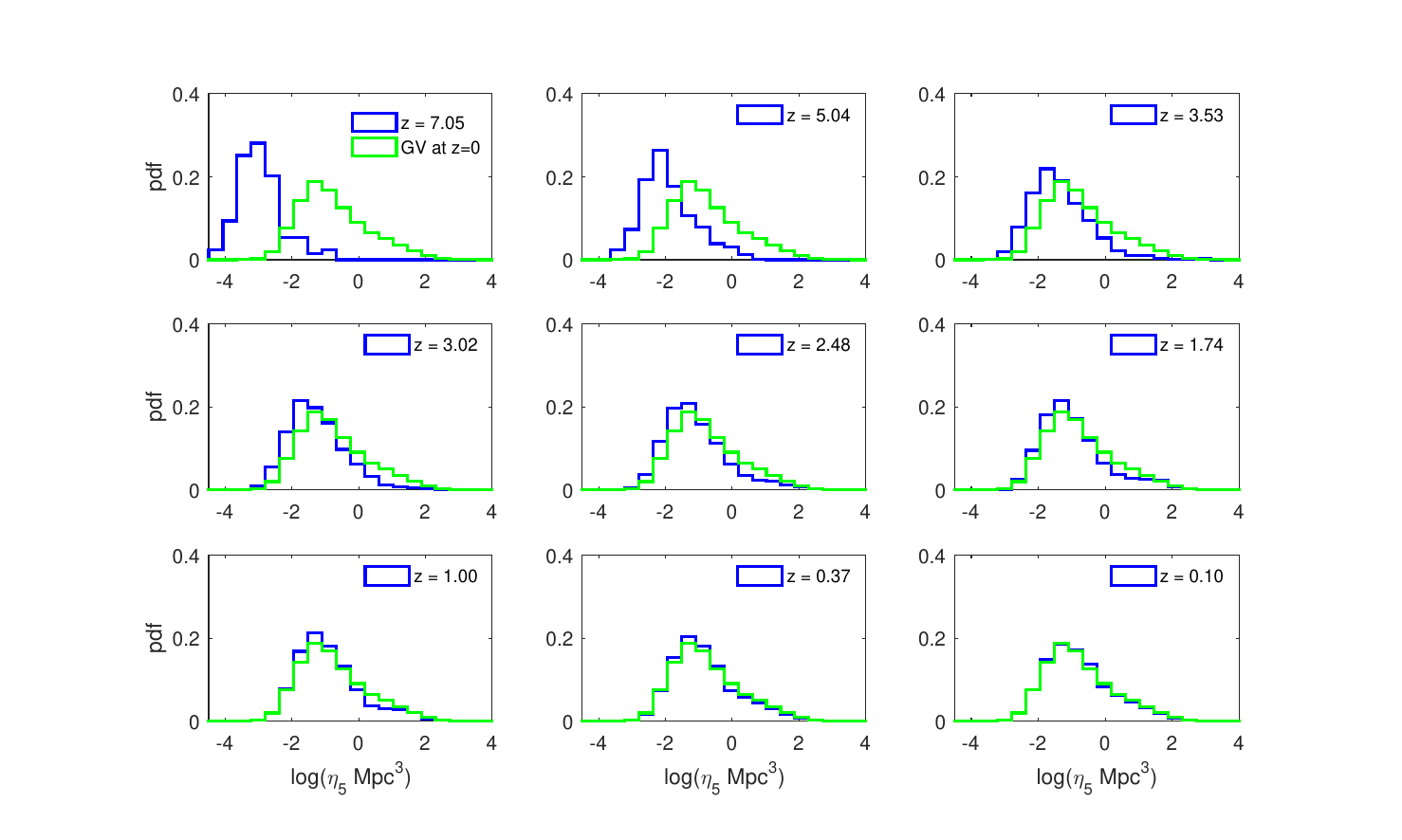}
\caption{This shows the PDF of the local density of the GV progenitors
  at various redshifts. The PDF of the local density of the
  present-day GV galaxies are shown together in each panel for
  comparison.}
\label{pdfden}
\end{figure*}

\begin{figure*}[htbp!]
%\centering
\includegraphics[width=17cm]{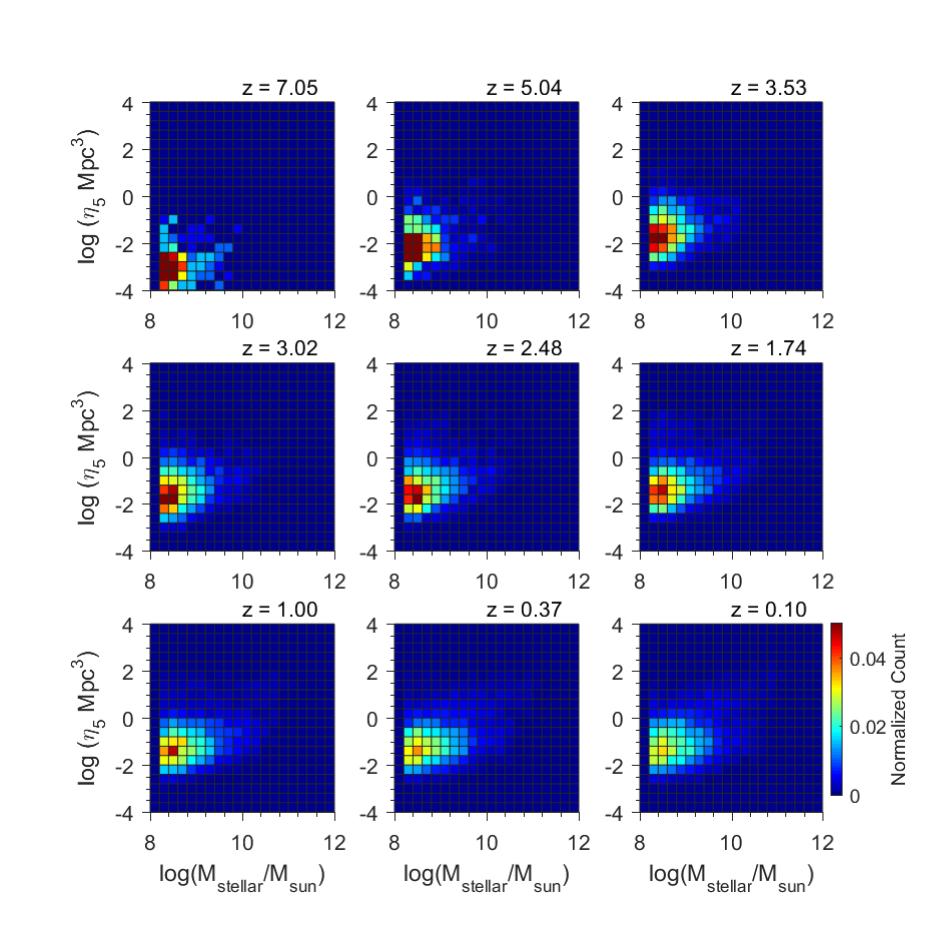}
\caption{This shows the joint PDF of stellar mass and local density of
  the GV progenitors at different redshifts. The colour bar represents
  the magnitude of the joint PDF of stellar mass and local density.}
\label{massden}
\end{figure*}

\begin{figure*}[htbp!]
\centering
\includegraphics[width=18cm]{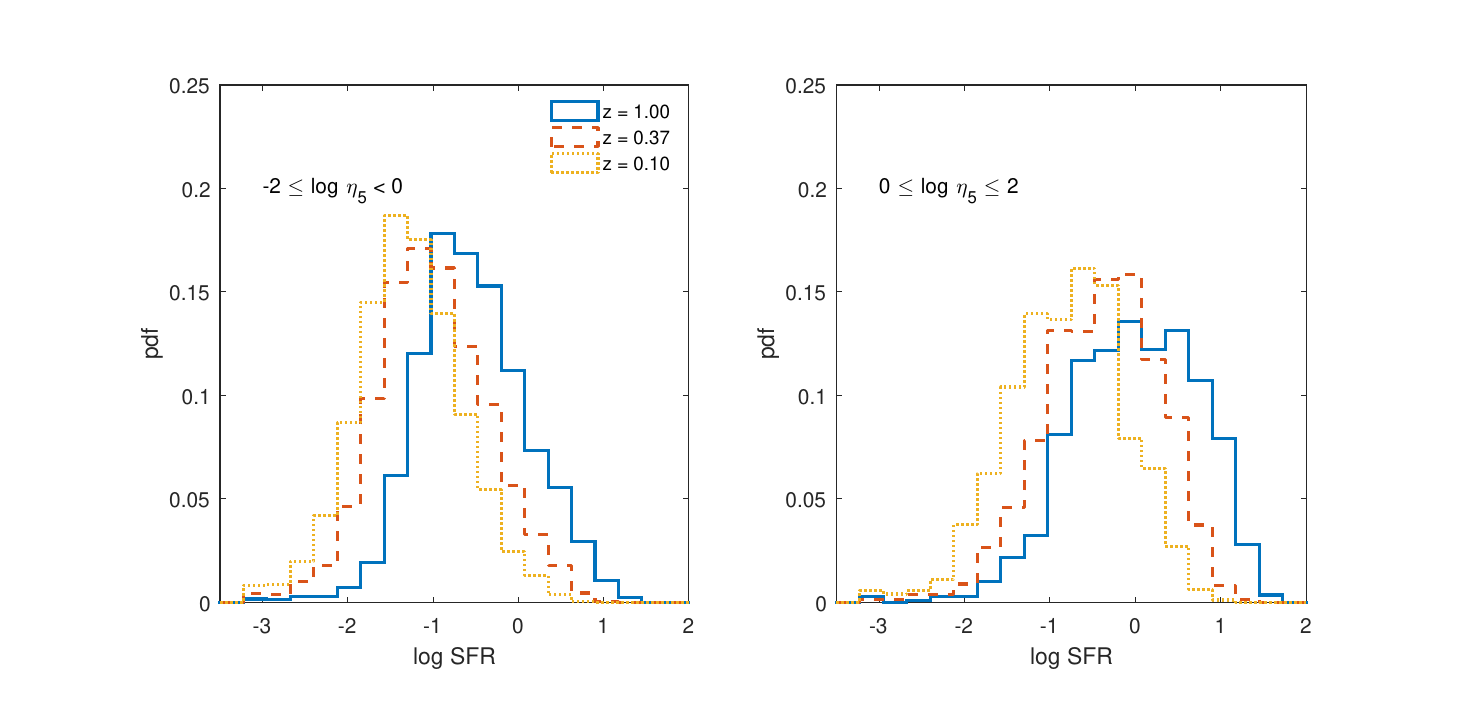}
\caption{The two panels of this figure show the PDF of the SFR for the
  GV progenitors at $z=0.1, 0.37, 1.00$ in low- and high- density
  regions.}
\label{densfr}
\end{figure*}

\begin{figure*}[htbp!]
\centering
\includegraphics[width=18cm]{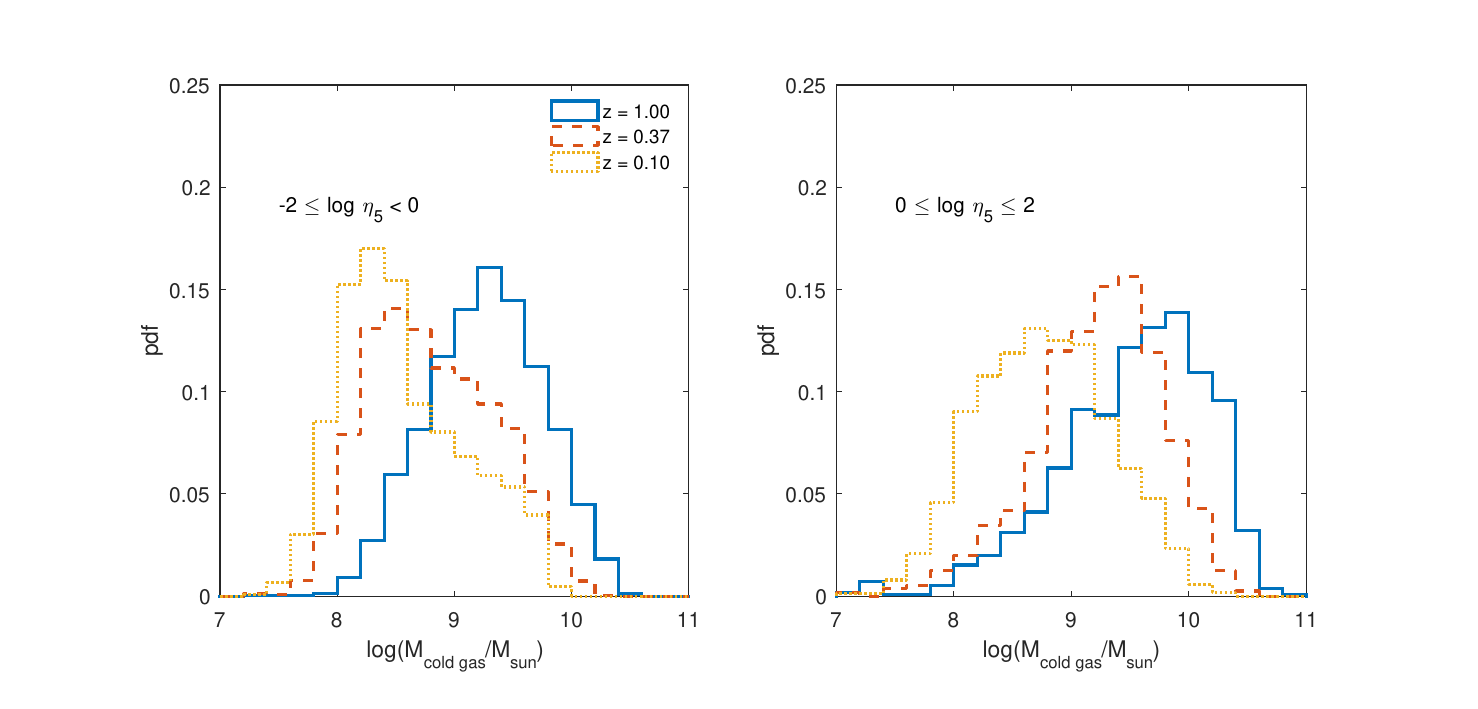}
\caption{The left panel of this figure shows the PDF of the cold gas
  content for the GV progenitors residing at low-density regions at
  $z=0.1, 0.37, 1.00$. The right panels shows the same but for the GV
  progenitors residing in the high density regions.}
\label{dencoldg}
\end{figure*}

\begin{figure*}[htbp!]
\centering
\includegraphics[width=18cm]{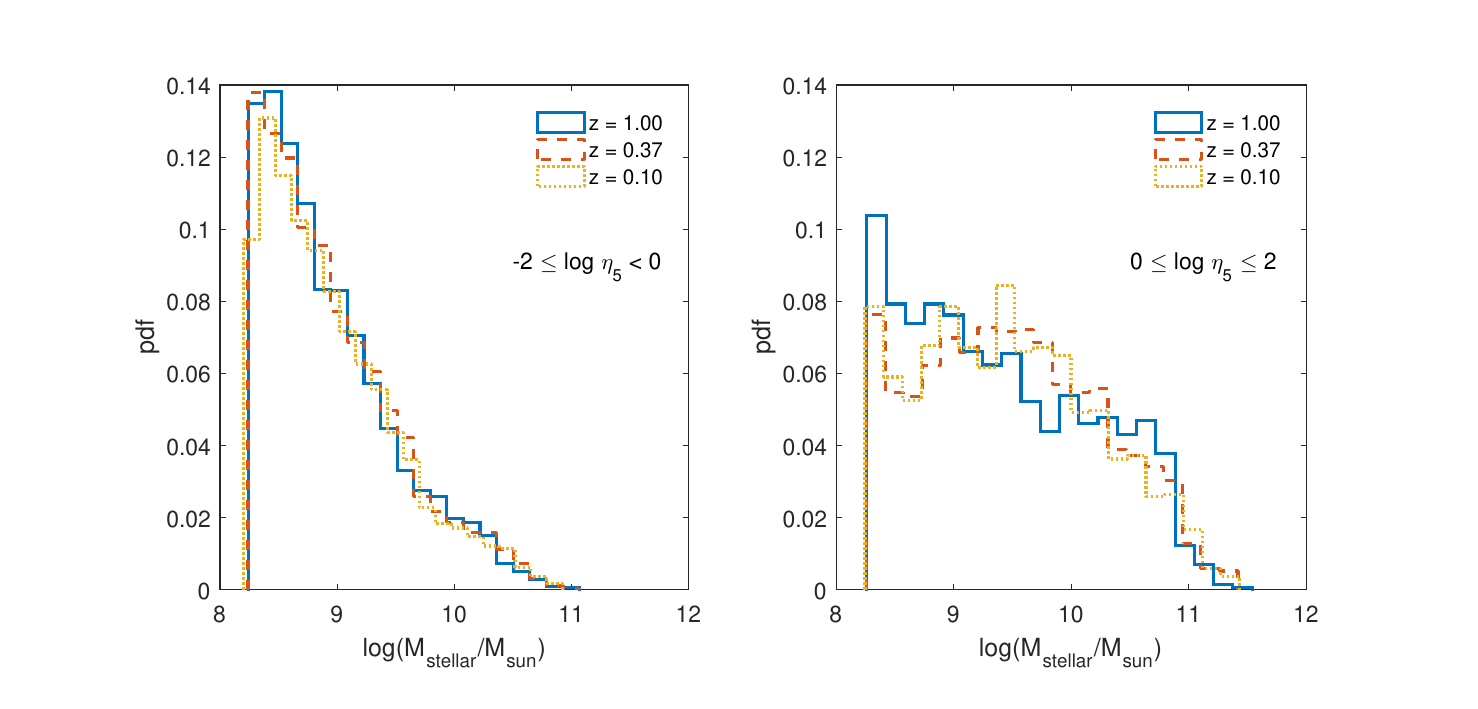}
\caption{The left panel and right panel of this figure respectively
  show the PDF of the stellar mass for the GV progenitors residing at
  low-density and high-density regions. The results are shown for
  three different redshifts $z=0.1, 0.37, 1.00$.}
\label{denm}
\end{figure*}

The mass of a galaxy's dark matter halo plays a pivotal role in
regulating its gas reservoir, which directly influences its star
formation potential. Massive halos, with their deeper gravitational
wells, can retain and accrete more gas from their surroundings while
better resisting feedback processes such as supernova-driven
outflows. These halos also facilitate the cooling and condensation of
hot gas into the cold phase necessary for star formation. In contrast,
low-mass halos are more vulnerable to gas loss due to feedback and
environmental effects, resulting in smaller gas reservoirs and reduced
star formation rates. This intricate interplay between dark matter
halo mass and gas dynamics is key to understanding the regulation of
star formation, the evolution of galaxies, and the patterns observed
in the star-forming main sequence.

The panels in \autoref{coldhalo} illustrate the joint PDFs of $\log
M_{cold\,gas}$ and $\log M_{halo}$ for GV progenitors across various
redshifts, as indicated. These panels reveal a strong correlation
between cold gas mass and halo mass among GV progenitors. At high
redshifts, this correlation is less pronounced, suggesting that cold
gas availability in GV progenitors is less tightly regulated by halo
mass. This trend likely reflects efficient cold gas accretion from the
cosmic web, enabling sustained star formation across a wide range of
halo masses.

At lower redshifts ($z<1$), the correlation steepens significantly,
indicating that halo mass becomes a stronger determinant of cold gas
availability. This shift may result from the increasing influence of
environmental quenching mechanisms at lower redshifts. The cold gas
loss from low-mass halos due to feedback or environmental interactions
can lead to rapid gas depletion and early quenching of star formation.
On the other hand, the high-mass halos may continue to accumulate gas
but often experience quenching through mechanisms like virial shock
heating or AGN feedback, which prevent the cooling and condensation of
gas necessary for star formation. The interplay between halo mass and
cold gas dynamics underscores the complex regulation of star formation
in GV progenitors.

The slope of the cold gas mass–halo mass relation for GV progenitors
at various redshifts is determined through least-squares fitting, as
shown in \autoref{coldhalo_slope}. This slope evolves notably with
redshift, decreasing from $z=10$ to $z=6$, then increasing slightly
between $z=6$ and $z=2$, and rising more significantly at $z<2$.

At $z>6$, the weak dependence of cold gas mass on halo mass suggests
that GV progenitors are in an early growth phase. During this time,
cold gas availability is abundant, with accretion largely unhindered
by halo mass constraints, fueling high star formation rates. The
reduced correlation may also reflect the regulatory effects of
significant AGN activity (\autoref{fracagn}), which can limit gas
accretion, particularly in massive halos.

Between $z=6$ and $z=2$, the dependency of cold gas mass on halo mass
strengthens as interactions and feedback mechanisms increasingly
regulate cold gas reservoirs. During this period, galaxy interactions
peak, with about $\sim 25\%$ of GV progenitors experiencing
interactions at $z=2$, triggering starbursts and accelerating cold gas
depletion. High-mass progenitors, however, manage to maintain their
cold gas more effectively, allowing them to sustain star formation for
longer periods. This marks a transitional era where halo properties
begin to play a more prominent role in shaping cold gas availability.

At $z<1$, cold gas reservoirs in GV progenitors deplete rapidly
(\autoref{cold}). Progenitors in low-mass halos are particularly
susceptible to feedback effects like supernova-driven outflows and
environmental quenching mechanisms such as ram pressure stripping,
leading to rapid gas loss and quenching of star formation. In
contrast, high-mass halos, with their deeper gravitational wells,
better resist gas stripping and retain cold gas for longer
durations. As GV progenitors increasingly migrate into intermediate-
and high-density regions, environmental processes selectively strip
cold gas from smaller halos, steepening the slope of the relation.

The sharp increase in slope at $z<1$ reflects a divergence in
progenitor evolution. The low-mass halos quench quickly, while
high-mass halos continue to sustain star formation during this phase,
highlighting the critical role of halo mass in late-stage galaxy
evolution.

\subsection{Roles of mass and local environment in the evolution of the progenitors of the present-day green valley galaxies}
We examine the evolution of the stellar mass distribution for the
progenitors of present-day GV galaxies in \autoref{pdfmass}. Each
panel compares the stellar mass distribution of GV progenitors at
different redshifts with that of present-day GV galaxies. The results
reveal that the high-mass end of the stellar mass distribution
gradually extends to larger values with decreasing redshift,
reflecting the hierarchical growth of structures. This growth is
driven by successive mergers of low-mass GV progenitors, which combine
to form more massive galaxies over time. At lower redshifts ($z<1$),
stellar mass growth continues at a slower pace.

We compare the local density distributions of GV progenitors with
present-day GV galaxies across different redshifts in
\autoref{pdfden}. The results show that GV progenitors gradually
transition from low-density regions at high redshifts to moderate and
high-density environments as redshift decreases. This shift becomes
particularly pronounced at $z<1$, where the migration into
intermediate- and high-density regions exposes these galaxies to
environmental quenching mechanisms. Despite this trend, the majority
of GV progenitors still reside in low- to intermediate-density
environments, even at lower redshifts. Additionally, the environmental
conditions of GV progenitors exhibit only moderate changes at $z<1$,
suggesting that while environment plays an increasing role in their
evolution, many GV progenitors remain in less dense regions where
quenching is less aggressive.

The joint PDF of stellar mass and local density for GV progenitors at
various redshifts is presented in \autoref{massden}. The results again
reveal a gradual migration of GV progenitors from low-density regions
to intermediate- and high-density environments as redshift
decreases. Until $z=3$, the majority of GV progenitors are situated in
low-density regions. Over time, as redshift decreases, these galaxies
gradually migrate to higher-density environments, highlighting the
role of environmental factors in their evolution. By $z<1$, the more
massive GV progenitors (with stellar masses exceeding $10^{10} \,
M_{sun}$) are predominantly found in high-density environments where
dynamical interactions and gas stripping suppress star formation,
leading to a faster progression through the green valley
phase. Despite this trend, most GV progenitors continue to reside in
low- to intermediate-density regions and have stellar mass $<10^{10}
\, M_{sun}$.

In \autoref{densfr}, \autoref{dencoldg} and \autoref{denm}, we analyze
the evolution of SFR, cold gas content, and stellar mass in GV
progenitors residing in low- and high-density environments. The two
panels of \autoref{densfr} compare the PDFs of SFR for GV progenitors
at $z=0.1, 0.37,$ and $1$ in low- and high-density regions,
respectively. Similarly, the panels of \autoref{dencoldg} present the
cold gas distributions in GV progenitors for these environments. These
figures reveal significant evolution in both SFR and cold gas
reservoirs for GV progenitors at $z<1$, regardless of their
environments. A clear reduction in cold gas content and a suppression
of SFR is observed with decreasing redshift. However, the PDFs show a
more pronounced shift in high-density environments, indicating faster
cold gas depletion and more rapid suppression of star formation in
these regions. This provides important insights into the evolutionary
pathways of GV progenitors. Low-mass progenitors evolve slowly in
low-density regions, with their development shaped by gradual
processes and slow gas depletion (\autoref{denm}). High-mass
progenitors in low-density environments also experience extended
periods in the green valley, primarily driven by internal processes.
In contrast, low-mass progenitors in high-density environments undergo
rapid quenching due to external environmental factors whereas the
high-mass progenitors hasten their entry into the green valley due to
their capability of retaining cold gas for a longer duration
(\autoref{denm}).

\begin{figure*}[htbp!]
\centering \includegraphics[width=15cm]{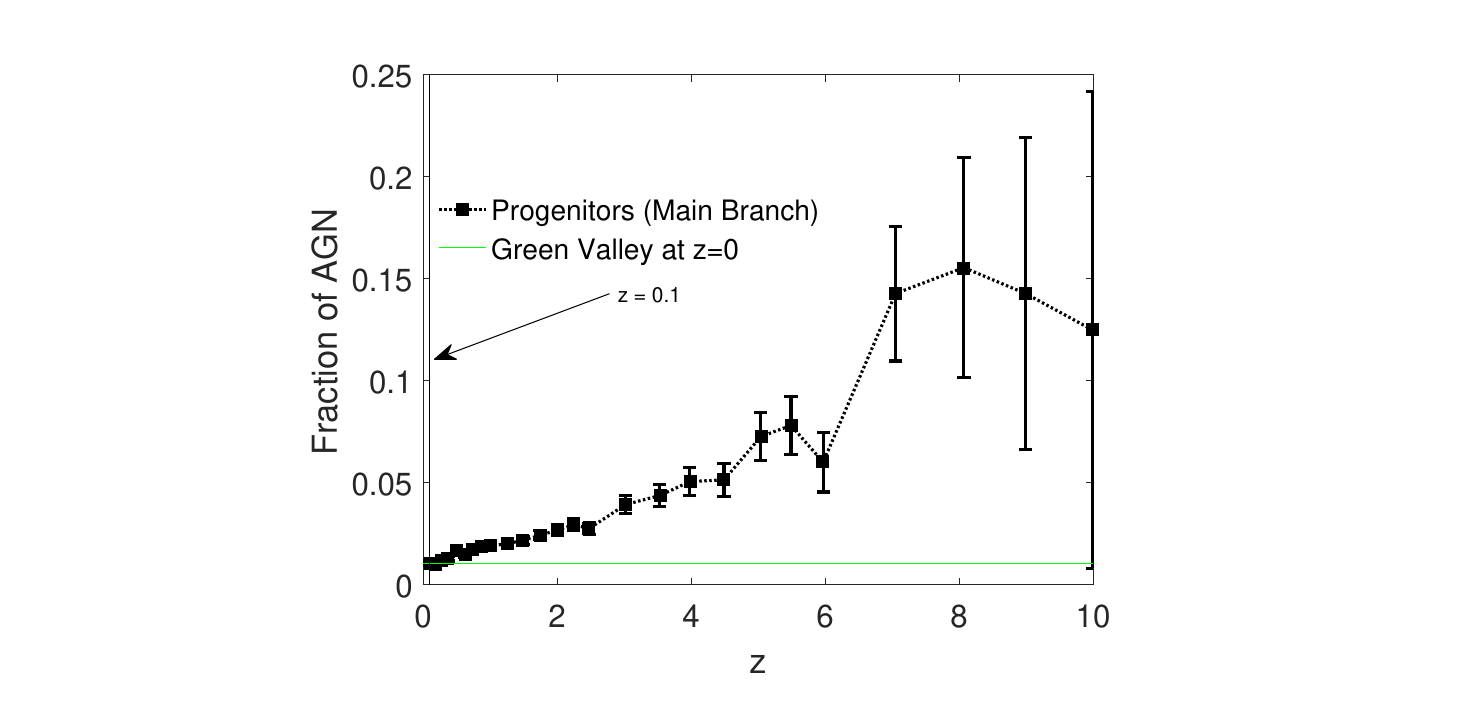}
\caption{Same as \autoref{fracagn} but only for the main progenitor
  branches of the $z=0$ GV galaxies.}
\label{agnmain}
\end{figure*}

\begin{figure*}[htbp!]
\centering \includegraphics[width=15cm]{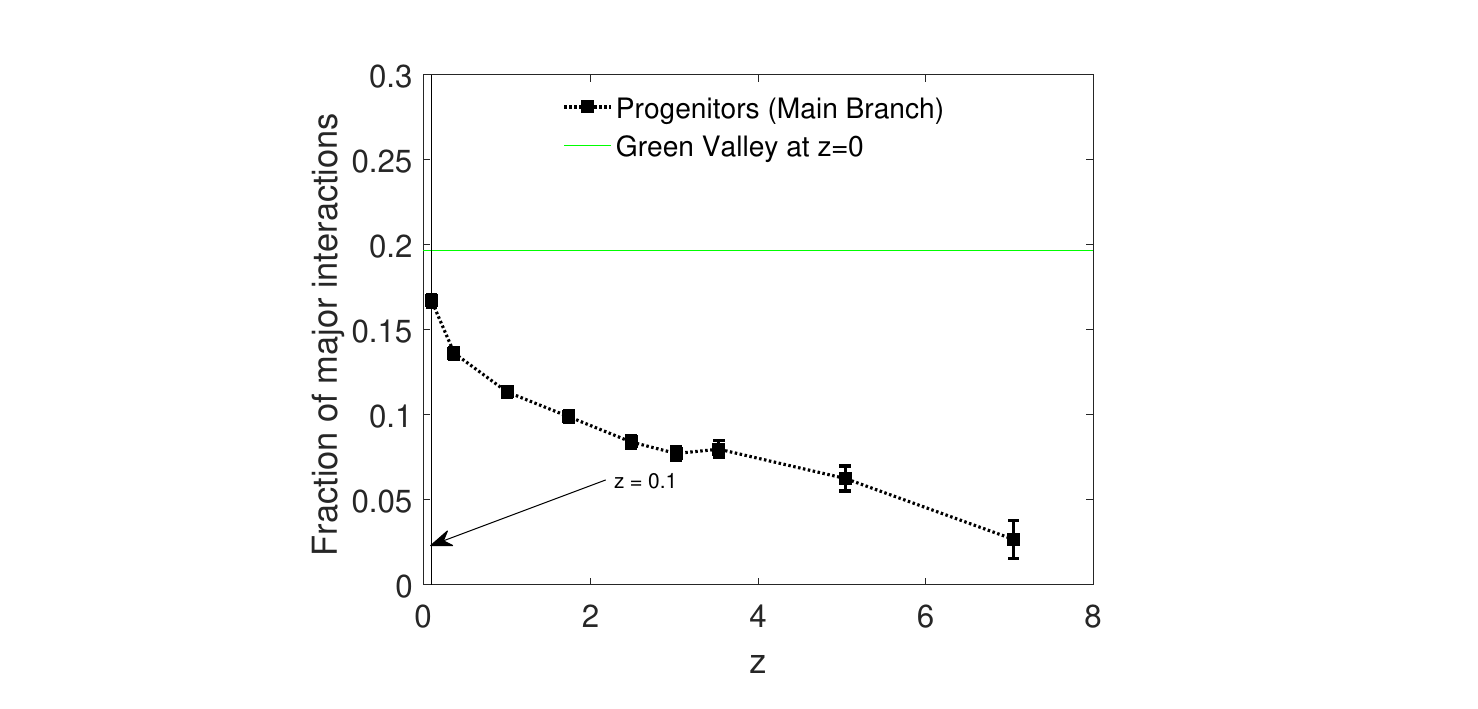}
\caption{This shows the fraction of main progenitor branches of $z=0$
  GV galaxies in major pairs at different redshifts.}
\label{intermain}
\end{figure*}

\begin{figure*}[htbp!]
\centering
\includegraphics[width=18cm]{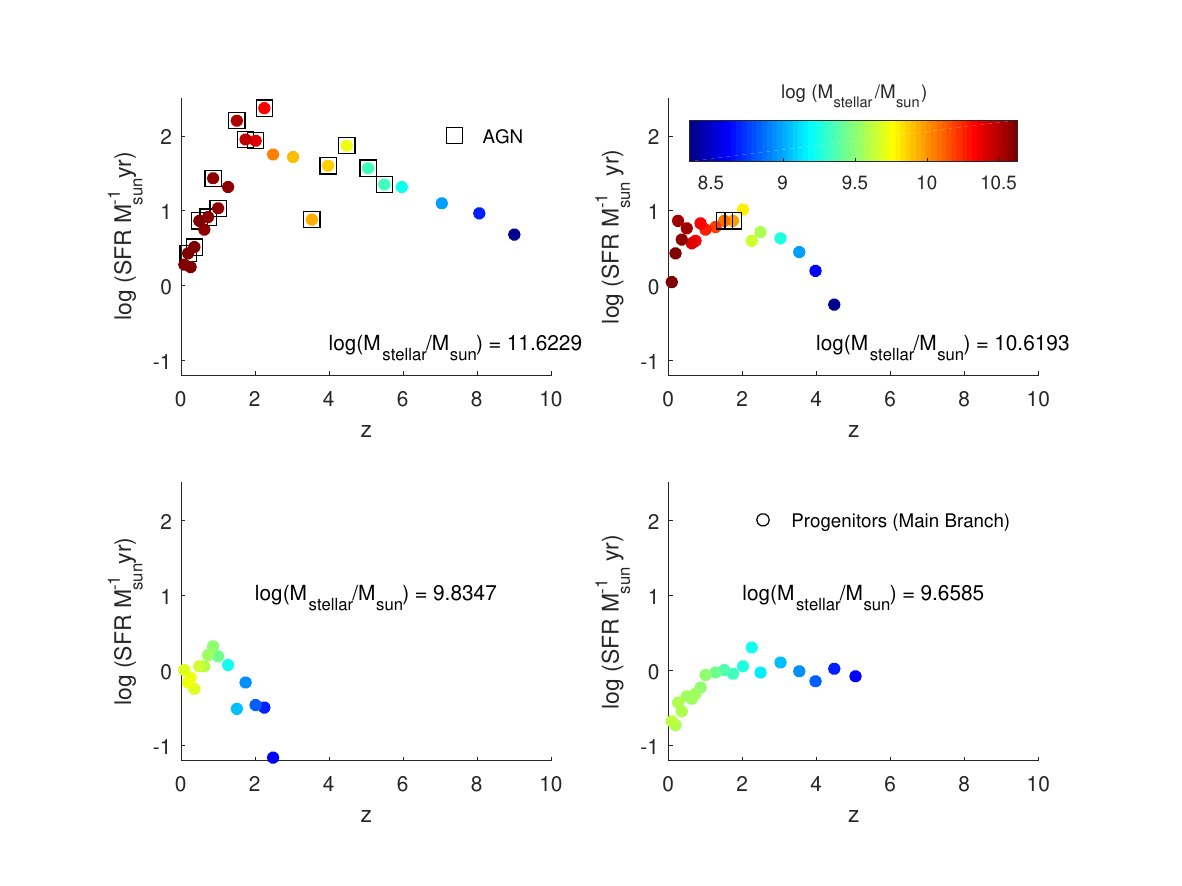}
\caption{This figure shows the evolution of SFR and stellar mass in
  main progenitor branches for 4 present-day GV galaxies. The present
  stellar mass of these four $z=0$ GV galaxies are mentioned in the
  respective panel. We mark the presence of AGN activity in the main
  progenitor branch at different redshift using rectengular boxes.}
\label{mainbranch1}
\end{figure*}

\begin{figure*}[htbp!]
\centering
\includegraphics[width=18cm]{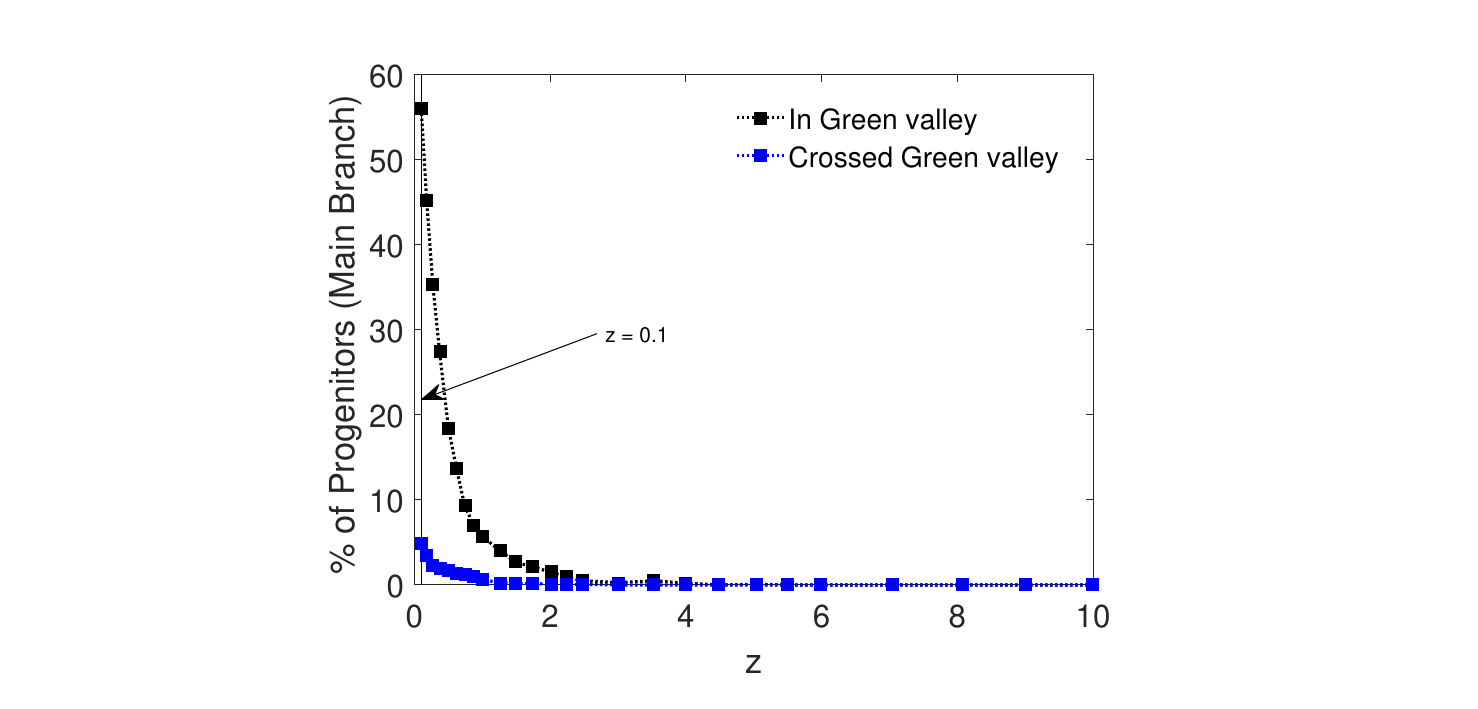}
\caption{This figure shows the fraction of the main progenitor
  branches of the present day GV galaxies that entered or crossed the
  green valley at different redshift. It is interesting to note that a
  small fraction ($\sim 5\%$) of the main progenitor branches has
  already crossed the green valley and entered the red sequence by
  $z=0.1$. These galaxies must have gone through some rejuvenation
  after $z=0.1$ that helped them to reenter the present-day green
  valley.}
\label{mainbranch2}
\end{figure*}

\subsection{Evolution of the main progenitor branches of the present-day green valley galaxies}
our sample of GV progenitors may include a significant number of
satellite galaxies which may bias the results in our analysis. In
order to asses their influence, we also repeat some of our analysis
considering only the main progenitor branches of the present day GV
galaxies. We show the evolution of the AGN fraction in the main
progenitor branches in \autoref{agnmain}. It shows that the evolution
of the AGN fraction in the main progenitor branches is quite similar
to \autoref{fracagn}. The similarities in \autoref{fracagn} and
\autoref{agnmain} suggest that our primary sample of GV progenitors is
not dominated by the satellite population. We note that a large number
of satellites are excluded by a cut in the stellar mass
($\log(M_{stellar}/M_{sun}) \geq 8.3$) of our primary sample of GV
progenitors. However, such a cut in the stellar mass would not exclude
all the satellite galaxies. A slightly higher magnitude of the AGN
fraction is observed in the main progenitor branches compared to the
primary sample of GV progenitors at all redshifts $z>2$
(\autoref{fracagn} and \autoref{agnmain}). This indicates that the AGN
fraction in the GV progenitors are somewhat underestimated at higher
redshifts due to the inclusion of the satellite population. Both
\autoref{fracagn} and \autoref{agnmain} show that AGN activity began
to wane at intermediate redshifts ($2<z<6$) and diminish significantly
at $z<2$.

We also calculate the fraction of main progenitor branches in major
pairs as a function of redshift in \autoref{intermain}. Interestingly,
we do not find a declining trend in the fraction of main progenitor
branches in major pairs at $z<2$. In fact, the fraction of main
progenitor branches in major pairs increases from $\sim 8\%-9\%$ at
$z=2$ to $\sim 17 \%$ at $z=0.1$. This emphasizes the role of
interactions in the evolution of the main progenitor branch even at
smaller redshift.

We show the evolution of SFR and stellar mass in main progenitor
branches for four $z=0$ GV galaxies in \autoref{mainbranch1}. The main
progenitor branches of two massive GV galaxies in the two upper panel
of this figure show steady rise in SFR until $z=2$. The results for
main progenitor branches of two less massive GV galaxies are shown in
the bottom two panels of this figure. We see that the SFR in the main
progenitor branch of the low mass GV galaxies increase or remain
unchanged with decreasing redshift before $z>1$.
\autoref{mainbranch1} shows that a dramatic decline in the SFR in the
main progenitor branch is observed at $z<1$ irrespective of the
stellar mass of the present-day GV galaxies. Both rapid mass growth
due to major mergers and onset of AGN activity likely play a
significant role in suppressing the formation in the main progenitor
branches of the two massive $z=0$ GV galaxies. The main progenitor
branches of the two low mass GV galaxies have lower SFR compared to
the high mass GV galaxies. They experience a stronger environmental
quenching at $z<1$.

Finally, in \autoref{mainbranch2}, we present the fraction of the main
progenitor branches of present-day ($z=0$) GV galaxies that entered or
transitioned through the green valley at various redshifts. We find
that most of the main progenitor branches of GV galaxies ($\sim 50\%$)
entered the green valley at $z<1$. Notably, a small subset ($\sim
5\%$) of the main progenitor branches had already crossed the green
valley and joined the red sequence by $z=0.1$. This suggests that
these galaxies must have experienced some form of rejuvenation after
$z=0.1$, likely through processes such as gas accretion or minor
mergers, which reignited star formation and allowed them to reenter
the green valley by the present day. This highlights the dynamic and
non-linear evolutionary pathways that galaxies can follow during their
journey through the green valley.

\section{Conclusions and Discussions}
We investigate the evolution of the progenitors of the present-day GV
galaxies across the redshift range $z=10-0$ using data from the EAGLE
simulations. At high redshift, most GV progenitors have stellar masses
below $10^{10} \, M_{sun}$. They are gas-rich and actively forming
stars (\autoref{cold}, \autoref{sfr} and \autoref{pdfmass}). At $z>6$,
their growth is dominated by the smooth accretion of cold gas and less
influenced by interactions and mergers (\autoref{interact},
\autoref{intermain}). $6\%-12\%$ of the GV progenitors shows AGN
activity during this period (\autoref{fracagn}). The AGN feedback
during this time suppress star formation in these GV progenitors,
moving them from the blue cloud into the green valley. The main
progenitor branches of the GV galaxies exhibit a similar behaviour
with a slightly higher fraction of AGN during this period
(\autoref{agnmain}). AGN activity starts to decline at intermediate
redshifts ($2<z<6$) and drops significantly at $z<2$. Only $2\%-3\%$
GV progenitors exhibit AGN activity at $z=0.1$ (\autoref{fracagn},
\autoref{agnmain}).

GV progenitors gradually migrate from low-density to intermediate- and
high-density environments with decreasing redshift (\autoref{pdfden},
\autoref{massden}). In high-density regions, interactions and mergers
between low-mass progenitors drive the formation of higher-mass
progenitors ($>10^{10} \, M_{sun}$). These interactions can trigger
starbursts, depleting cold gas reservoirs, and accelerate their
transition into the green valley. The fraction of interacting
progenitors rises from $\sim 5\%$ at $z=7$ to $\sim 25\%$ at $z=2$
(\autoref{interact}), emphasizing the critical role of galaxy
interactions in shaping GV evolution. At $z<2$, the fraction of
interacting progenitors declines, indicating a diminished role for
these processes in quenching star formation at lower
redshifts. However, the fraction of the main progenitor branches of GV
galaxies keeps increasing even at $z<2$ (\autoref{intermain})
indicating the roles of interactions and mergers in their migration to
the green valley. Most of the main progenitor branches ($\sim 50\%$)
of the present-day GV galaxies eneter the Green valley after
$z<1$. Around $\sim 5\%$ of the main progenitor branches of the GV
galaxies cross the green valley to enter the red sequence by
$z=0.1$. These progenitors experience some kind of rejuvenation after
$z=0.1$ to reenter the green valley.

At $z>2$, the stellar mass of progenitors and their host dark matter
halos have a less decisive impact on star formation and
quenching. However, at $z<1$, tighter correlations between stellar
mass and SFR (\autoref{sfrmass}, \autoref{sfrmass_slope}), and halo
mass and cold gas (\autoref{coldhalo}, \autoref{coldhalo_slope})
content emerge , suggesting these factors become more influential. A
sharp decline in cold gas content and a rapid suppression of star
formation at $z<1$ (\autoref{cold}, \autoref{sfr}) highlight the
growing importance of environmental processes in the evolution of GV
progenitors. The evolution of GV progenitors is shaped by the
interplay between environment and stellar mass. Secular processes
dominate in low-density environments for lower-mass progenitors, while
environmental quenching mechanisms prevail in high-density regions for
massive progenitors. Once the universe entered a phase of accelerated
expansion at low redshift ($z<1$), interactions in low-density regions
likely became much less frequent.

In our analysis, the evolution of green valley (GV) progenitors can be
broadly divided into three distinct phases:\\

(i) Early growth phase ($z=10$ to $z=6$) during which GV progenitors
have abundant cold gas, enabling efficient gas accretion and high star
formation rates. They predominantly reside in low-density environments
with minimal external quenching influences. AGN feedback plays a key
regulatory role, particularly in massive progenitors, by moderating
star formation activity.\\

(ii) Transition phase ($z=6$ to $z=2$) during which GV progenitors
begin migrating toward intermediate- and high-density
regions. Frequent galaxy interactions in dense environments trigger
starbursts, drive AGN activity, and disrupt gas reservoirs. These
processes enhance the correlation between cold gas mass and halo
properties, as feedback and dynamical effects increasingly regulate
the gas content.\\

(iii) Quenching phase ($z=2$ to $z=0$) which is marked by strong cold
gas depletion and rapid suppression of star formation. Quenching
becomes increasingly governed by environmental and mass-dependent
processes, with high-density environments and internal mechanisms
playing critical roles in halting star formation.

In the quenching phase, the evolution of GV progenitors is shaped by
the interplay between mass and environmental density, leading to
diverse trajectories. The low-mass GV progenitors in low-density
environments evolve slowly, with moderate star formation sustained by
less effective environmental quenching and gradual depletion of cold
gas. The prolonged green valley phase observed in these galaxies may
result from the gradual influence of secular processes. The quenching
in high-mass GV progenitors in low-density environments could be
primarily governed by internal mechanisms, including AGN activity,
mass quenching, morphological quenching, and bar quenching. This
results in a prolonged residence in the green valley. On the other
hand, the low-mass GV progenitors in high-density environments would
experience rapid cold gas loss due to strong environmental quenching
mechanisms, such as ram pressure stripping and galaxy harassment,
leading to a quick transition through the green valley. High-mass GV
progenitors in dense environments are likely quenched by a combination
of internal feedback mechanisms (such as AGN feedback) and external
environmental influences. It may be noted that our analysis excludes
all the GV progenitors with $\log(M_{stellar}/M_{sun}) < 8.3$. Despite
this, we find that the majority of GV progenitors are lower-mass
galaxies ($<10^{10} \, M_{sun}$) residing in low- to
intermediate-density regions. As a result, their transition through
the green valley is generally prolonged, contributing to the extended
evolutionary journey of the present-day GV population.

It is important to discuss some caveats in our analysis. In this
analysis, we focus exclusively on GV progenitors with stellar masses
$\log(M_{stellar}/M_{sun}) \geq 8.3$ across the entire redshift
range. This exclusion of progenitors with masses below $8.3 \times
10^{8}\, M_{sun}$ introduces biases, particularly underestimating the
contributions of low-mass progenitors to the overall evolution of GV
galaxies. Low-mass progenitors, especially in low-density
environments, are likely underrepresented in this study. These
galaxies typically evolve more gradually, driven by secular processes
and slower gas depletion. Their omission may lead to an overemphasis
on rapid quenching mechanisms and interactions, skewing the
evolutionary picture toward higher-mass systems. Moreover, low-mass
progenitors in low-density environments often transition through the
green valley more slowly due to less efficient quenching, and their
exclusion could somewhat exaggerate the prevalence of rapid
transitions. Additionally, low-mass galaxies are more sensitive to
environmental effects such as ram pressure stripping and tidal
interactions, particularly in high-density regions. Excluding them
provides an incomplete understanding of how environmental quenching
operates across the full mass spectrum. These biases will be more
pronounced at higher redshifts, where progenitors tend to be less
massive.

Finally, we compare our findings with previous studies utilizing both
simulations and observations. Using COSMOS survey data, \cite{bundy10}
report evidence of mergers in green valley (GV) galaxies at $z \sim
1.2$. \cite{goncalves12} find that GV galaxies were preferentially
more massive in the past. \cite{salim14} propose a quasi-static
picture where most GV galaxies were partially quenched in the distant
past and are now experiencing a gradual decline in star
formation. \cite{schawinski14} identify multiple quenching mechanisms
in GV galaxies, including truncation of gas supply followed by gradual
depletion, and gas reservoir destruction during major mergers or AGN
feedback. Using EAGLE simulations, \cite{trayford16} investigate
galaxy colour evolution and find that most GV galaxies remain in the
green valley for less than 2 Gyr, regardless of the quenching
mechanism. \cite{coenda18} emphasize the environmental dependence of
external quenching mechanisms in GV galaxies. Analyzing data from the
Hyper Suprime-Cam Survey, \cite{jian20} observe a strong mass
dependence and mild environmental influence in the redshift evolution
of GV galaxies. In the nearby universe, \cite{das21} use SDSS data to
reveal that secular processes are more influential than
environment-driven quenching in suppressing star formation in
present-day GV galaxies. However, this analysis primarily focused on
the higher mass green valley galaxies. \cite{estrada23}, analyzing HST
data, find that GV progenitors were more compact than present-day
star-forming galaxies, indicating significant morphological
evolution. \cite{brambila23} study GV galaxies in the nearby universe
($z \leq 0.11$) and conclude that their evolution is governed by
distinct mechanisms depending on the environment. Our findings align
with several aspects of earlier studies while offering some nuanced
insights into the evolution of galaxies in the green valley.

In conclusion, our study indicates that GV progenitors undergo a
distinct evolutionary transition over cosmic time, shifting from
AGN-dominated quenching at high redshifts ($6 \leq z < 10$) to
interaction-driven evolution at intermediate redshifts ($2 \leq z <
6$), and finally to environmentally driven quenching at low redshifts
($0 \leq z < 2$). These processes, working in tandem with the
progressive depletion of cold gas, shape the transformation of the GV
population over cosmic time.

\section{Acknowledgements}
We thank an anonymous reviewer for the valuable comments and
suggestions that helped us to improve the draft. BP acknowledges the
support provided by IUCAA, Pune, through its associateship
program. The authors extend their gratitude to the Virgo Consortium
for providing public access to their simulation data. The EAGLE
simulations were performed using the DiRAC-2 facility at Durham,
managed by the ICC, and the PRACE facility Curie based in France at
TGCC, CEA, Bruy\`{e}res-le-Ch\^{a}tel.

\section{Data availability}
The EAGLE simulation data are publicly accessible at
https://icc.dur.ac.uk/Eagle/database.php. Data generated in this study
will be made available upon reasonable request to the authors.

\end{document}